\documentclass{article}

\usepackage{arxiv}

\usepackage[utf8]{inputenc} 
\usepackage[T1]{fontenc}    
\usepackage{textcomp}       
\usepackage{eurosym}        
\usepackage{url}            
\usepackage{booktabs}       
\usepackage{amsmath}
\usepackage{amsfonts}       
\usepackage{amssymb}
\usepackage{nicefrac}       
\usepackage{microtype}      
\usepackage{graphicx}
\usepackage[numbers,sort&compress]{natbib}
\usepackage[table]{xcolor}
\usepackage[ruled,vlined]{algorithm2e}
\usepackage{subcaption}
\usepackage{tabularx}
\usepackage{multirow}
\usepackage{makecell}
\usepackage{threeparttable}
\usepackage{siunitx}
\usepackage{comment}
\usepackage{tikz}
\usetikzlibrary{arrows.meta, positioning, shapes.geometric, fit, backgrounds, calc}
\usepackage{hyperref}       
\usepackage{doi}

\definecolor{bestgreen}{RGB}{226,239,218}
\definecolor{baselinegray}{RGB}{242,242,242}
\definecolor{dflblue}{RGB}{214,228,240}
\definecolor{mserow}{RGB}{242,242,242}
\definecolor{oraclegreen}{RGB}{198,224,180}

\title{Decision-Focused Learning for Optimal Battery Scheduling Under Load and Photovoltaic Power Uncertainty in Microgrids}

\author{Marios Saleptsis\thanks{Corresponding author.} \\
	Department of Energy\\
	Politecnico di Milano\\
	20156 Milano, Italy \\
	\texttt{marios.saleptsis@polimi.it} \\
	\And
	Antoine Lesage-Landry \\
	Polytechnique Montr\'eal, GERAD \& Mila\\
	Montr\'eal, QC H3T 1J4, Canada \\
	\texttt{antoine.lesage-landry@polymtl.ca} \\
	\AND
	Marco Mussetta \\
	Department of Energy\\
	Politecnico di Milano\\
	20156 Milano, Italy \\
	\texttt{marco.mussetta@polimi.it} \\
	\And
	Sonia Leva \\
	Department of Energy\\
	Politecnico di Milano\\
	20156 Milano, Italy \\
	\texttt{sonia.leva@polimi.it} \\
}

\renewcommand{\shorttitle}{Decision-Focused Learning for BESS Scheduling in Microgrids}

\hypersetup{
pdftitle={Decision-Focused Learning for Optimal Battery Scheduling Under Load and Photovoltaic Power Uncertainty in Microgrids},
pdfsubject={eess.SY, cs.LG},
pdfauthor={Marios Saleptsis, Antoine Lesage-Landry, Marco Mussetta, Sonia Leva},
colorlinks=true, linkcolor=blue!60!black, citecolor=blue!60!black, urlcolor=blue!60!black,
pdfkeywords={Decision-focused learning, End-to-end learning, Microgrid, Forecasting, BESS scheduling, Photovoltaic, Load demand},
}

\begin{document}
\maketitle

\begin{abstract}
The increasing penetration of renewable energy sources in microgrids demands accurate day-ahead forecasting to support optimal battery energy storage system (BESS) scheduling. Yet the conventional predict-then-optimize paradigm trains forecasters on statistical accuracy metrics that are agnostic to the asymmetric and time-varying cost impacts of forecast errors. This work proposes a decision-focused learning (DFL) framework that trains gated recurrent unit forecasters for both load demand and photovoltaic generation directly on the downstream operational cost of a two-stage energy management system, where day-ahead BESS commitments are fixed prior to real-time balancing. Gradients are backpropagated from the real-time operational cost through both optimization stages to the forecaster weights via implicit differentiation, enabling the forecasters to internalize the penalty structure of the scheduling problem. The framework is validated on the MG\textsuperscript{2}Lab experimental microgrid testbed at Politecnico di Milano across two operational scenarios of increasing grid constraint and eight forecasting configurations comprising single-source and combined-forecaster settings. Results illustrate that the DFL-trained pipeline reduces operational cost by 3.78\% under normal grid conditions and by 30.60\% under grid-constrained conditions relative to the accuracy-optimized baseline, while nearly eliminating load shedding events. Notably, the DFL forecasters exhibit higher statistical errors than their accuracy-trained counterparts yet achieve consistently lower operational costs, confirming that forecast accuracy and decision quality do not share a monotonic relationship in constrained microgrid scheduling.
\end{abstract}


\section{Introduction}
\label{sec:intro}
The European Union's commitment to achieving climate neutrality by 2050, underpinned by intermediate targets of 55\% greenhouse gas emissions reduction by 2030, necessitates unprecedented integration of renewable energy sources into industrial power systems \cite{zero_emission, nature_2}. 
However, the inherent intermittency and uncertainty of renewable generation pose significant challenges for real-time energy management and operational planning \cite{XGBRef}. 

Accurate short-term forecasting has become mandatory for microgrid operators seeking to maintain supply-demand balance, minimize operational costs, and maximize the utilization of low emission generation sources \cite{load_forecast,ems_2}.
In microgrid (MG) energy management systems (EMSs), day-ahead forecasts serve as core inputs to optimization algorithms responsible for unit commitment, battery charge-discharge scheduling, and grid power exchange decisions \cite{OptMethods, ems_3}. The traditional approach follows the predict-then-optimize (PTO) paradigm, wherein forecasts are first generated independently, then subsequently utilized as deterministic parameters in operational optimization problems \cite{PTO_1}. 
Contrastingly, emerging decision-focused learning (DFL) frameworks propose end-to-end (E2E) optimization that directly minimizes decision costs rather than forecast errors \cite{Donti_NRIPS}. 
For example, this can be done through the utilization of differentiable layers in the training process of the forecasting module \cite{gradient_based_intro}, or derivative-free techniques associating the forecasting optimization layer with the downstream operational cost of the MG, creating a black-box automated pipeline \cite{derivative_free}. 

Regardless of the optimization paradigm employed, forecast errors can propagate directly into suboptimal operational decisions, resulting in incorrect scheduling and unit commitment, and eventually in  excessive electricity procurement costs \cite{mike_forecasting, two_stage_els}.
Despite this clear operational impact, the current evaluation methodology in forecasting literature focuses mainly on statistical accuracy metrics, without validating whether incremental improvements in these metrics translate proportionally to reductions in actual operational costs. This disconnect between forecast accuracy and economic performance represents a critical gap in the assessment of the true efficiency of forecasting models in a operational context.



\section{Literature Review}
\label{sec:lit}

This section provides a structured literature review spanning the PTO and DFL paradigms as applied to energy system scheduling and forecasting.

\subsection{Predict-then-Optimize in Energy Systems}
The prevailing approach to microgrid energy management follows the PTO paradigm, in which forecasting models for photovoltaic (PV) generation, load demand, and electricity prices are trained independently of the downstream scheduling problem. Under this paradigm, point or probabilistic predictions are generated first and subsequently fed as fixed parameters into the optimization problem that determines optimal dispatch decisions \cite{Parisio2014_TCST, Parisio2014_ApplEnergy}. Two-stage stochastic or robust extensions further handle forecast uncertainty through scenario generation or worst-case budgets, yet still treat the forecaster and the optimizer as decoupled modules \cite{TwoStageStochastic_MG, RobustMG_DayAhead}. Although these methods have proven effective for real-time cost reduction and constraint satisfaction, they inherently assume that minimizing the statistical forecast error is the best proxy for minimizing the operational cost of the microgrid.

However, a growing body of evidence challenges this assumption. Several studies have shown that the relationship between forecast quality and operational value is not monotonic: a model with lower RMSE does not necessarily yield lower scheduling cost \cite{trueforecast, mike_forecasting}. In the context of battery energy storage arbitrage, recent work demonstrated that conventional accuracy metrics such as RMSE and MAE are only weakly correlated with the realized profit, whereas measures capturing the alignment of predicted and actual price curves exhibit a stronger link to economic performance \cite{staterror}. Similarly, in local energy communities, the so-called ``value'' of a forecast, measured through operational indicators like the total operational cost of and supply cover factor, can diverge from the ranking produced by standard statistical metrics \cite{forecast_value}. This disconnect arises because statistical loss functions weight all errors uniformly regardless of their location in the decision space, whereas in practice, errors at times of high marginal cost or near constraint boundaries disproportionately affect the operational outcome. The recognition of this accuracy-to-cost mismatch has motivated the development of decision-aware training paradigms.

\subsection{Decision-focused Learning in Energy Systems}
DFL, also termed end-to-end or task-based learning, trains a forecasting model directly on the downstream optimization objective rather than on a statistical loss. Its methodological foundations are surveyed in \cite{Mandi2024_DFL_review}, who organize gradient-based techniques into four families spanning analytical differentiation, smoothing, random perturbation, and surrogate losses, and in \cite{Kotary2021_E2E_survey} for the broader integration of solvers with learning. Within energy systems, these ideas have been applied at progressively finer scales, each step moving closer to asset-level scheduling while leaving distinct pieces of the microgrid problem unaddressed.

At the system level, \cite{DFL_ellinas_EPSR} shows that deliberately biasing a demand forecast can steer day-ahead unit commitment toward lower system cost, though their Bayesian-optimization approach scales poorly and depends on a synthetic price model. Narrowing the scope to a single building, \cite{DFL_HVAC_smartgrids} embeds a neural network as exact constraints inside a cooling system optimization to learn through a non-differentiable simulator, yet omitted storage, renewables, and any market interaction.

Bringing storage and markets into view, \cite{Stratigakos2022_PrescriptiveTrees} employs prescriptive trees for renewable trading and arbitrage as an interpretable alternative, and \cite{DFL_reservemarket_segan} couples a price regressor to a differentiable optimization layer for BESS reserve-market participation; both, however, operate in wholesale or reserve settings with no load or PV uncertainty. Reference \cite{perturbed_DFL_IEEE} pushes single-asset arbitrage further with perturbation-based differentiation, while \cite{Beichter2024_DFL_Retraining} targets the recurring scalability bottleneck through decision-focused retraining of a pre-trained forecaster. These advances sharpen DFL for storage, but always around a single uncertain input and outside any microgrid.

The residential and microgrid setting is where DFL meets battery energy storage system (BESS) scheduling most directly, and here the remaining gaps come into focus. The authors of \cite{DFL_featureeng_expertsystems} couple automated feature engineering with decision-focused learning for energy-storage scheduling, improving operational cost over accuracy-trained forecasters, whereas \cite{Huang2025_DataCentric_TOL} refines the training data itself for PV-BESS operation while treating load as known. Closest to the present study, \cite{DFL_KULeuven} trains a long short-term memory PV forecaster on the cost of operating a PV-battery system under dynamic pricing, yet still assumed known load, omitted grid constraints, and relied on a simulated battery with synthetically perturbed irradiance rather than real forecasts.

\subsection{Research Gaps and Contributions}
Despite growing interest in DFL for energy systems, the reviewed literature reveals four gaps. First, prior DFL studies in the BESS domain address a single source of uncertainty, leaving the interaction of multiple sources within a unified EMS unexplored. Second, although several works report that DFL forecasters incur higher statistical error yet lower operational cost, this non-monotonic relationship between forecast accuracy and economic performance has been neither systematically quantified nor mechanistically explained. Third, no study examines how the microgrid's operational flexibility governs the size of the DFL advantage over PTO. Finally, existing frameworks have been validated almost exclusively in simulation, using synthetic benchmarks, residential-scale systems, or simulated batteries, with none demonstrated on an experimental industrial microgrid that combines real PV generation, a physical BESS, and realistic grid-exchange constraints.

These gaps collectively motivate the present work, whose contributions are outlined as follows.
\begin{itemize}
    \item We propose the first application of decision-focused learning that simultaneously addresses both load demand and PV generation uncertainty in a microgrid EMS, extending DFL beyond the single-source settings considered in prior work by training independent decision-focused forecasters for each uncertainty source.

    \item We propose a differentiable two-stage optimization architecture in which day-ahead BESS commitments are fixed prior to real-time operation, ensuring that forecast errors propagate into real operational costs through explicit load shedding penalties, thereby providing a stronger and more directional gradient signal than symmetric accuracy-based losses.

    \item We validate the proposed framework on the MG\textsuperscript{2}Lab  industrial microgrid testbed under two operational scenarios of increasing grid constraints, enabling a controlled assessment of how grid flexibility governs the value of decision-focused forecasting relative to the conventional PTO paradigm.

    \item We design a comprehensive factorial evaluation framework that serves as an empirical roadmap for decision-focused forecasting in microgrids: eight configurations isolate the marginal contribution of DFL for load and PV independently and pair both training methodologies across the two sources, enabling a rigorous end-to-end comparison of decision-focused and accuracy-driven forecasting.
\end{itemize}

\section{Problem Formulation}
\label{text3:problem}

The MG energy management problem addressed in this work consists of scheduling a BESS over a day-ahead horizon of $T$ timesteps with temporal resolution $\Delta t$ to minimize operational cost, given uncertain load demand and PV generation. We denote the set of scheduling timesteps by $\mathcal{T} = \{0, 1, \dots, T-1\}$. The core challenge lies in the sequential nature of the decision process: the BESS commitment must be made in advance from forecasts of both load and PV generation, yet its cost consequences are only realized at real time, once actual demand and generation are revealed. This temporal asymmetry motivates the two-stage formulation adopted here, in which day-ahead decisions are optimized under forecast uncertainty and real-time balancing is achieved through grid exchange and, if necessary, load shedding. The following subsections describe the general MG structure and the formulation of the two-stage EMS.

\subsection{System Model}
\label{text3:mg}

The microgrid considered in this work comprises three main components: a PV generation unit with rated capacity $P_{\text{pv},\text{max}}$, a BESS, and a stochastic load that includes electric vehicle (EV) charging with peak demand $P_{\text{ev},\text{max}}$, all interfaced with the main distribution grid through a bidirectional connection.

The PV plant represents the primary renewable generation source, and its output $p_{\text{pv},t}$ is treated as a source of uncertainty in the EMS formulation, with day-ahead forecast denoted by $\hat{p}_{\text{pv},t}$. The load introduces a stochastic, time-varying demand component $y_t$, with day-ahead forecast $\hat{y}_t$; the presence of EV charging contributes significant variability and makes accurate day-ahead load forecasting a critical operational challenge. Both quantities are forecast independently  and enter the day-ahead optimization as the two uncertain inputs.

The BESS is the primary controllable asset of the microgrid and constitutes the central decision variable of the EMS. Its usable energy window is bounded between a minimum state of energy $SOE_{\text{min}}$ and a maximum $SOE_{\text{max}}$ to prevent deep discharge and overcharge, preserving battery longevity. The system is initialized at $SOE_{\text{init}}$ at the beginning of each scheduling horizon. Charge and discharge power are bounded by rated limits $P_{\text{ch},\text{max}}$ and $P_{\text{dis},\text{max}}$, with charge and discharge efficiencies $\eta_{\text{ch}}$ and $\eta_{\text{dis}}$, respectively. An operation and maintenance (O\&M) cost $c_{\text{OM}}$ (in \texteuro/kWh) is applied to all discharge throughput to account for battery degradation.

Two penalty costs govern the economic formulation of the EMS. PV curtailment is penalized at a rate $c_{\text{curt}}$ to discourage unnecessary wastage of renewable generation, while load shedding carries a penalty $c_{\text{shed}}$, set to reflect the true cost of unmet demand. Grid exchange is priced through time-varying purchase and sell tariffs $\lambda_{\text{buy},t}$ and $\lambda_{\text{sell},t}$.

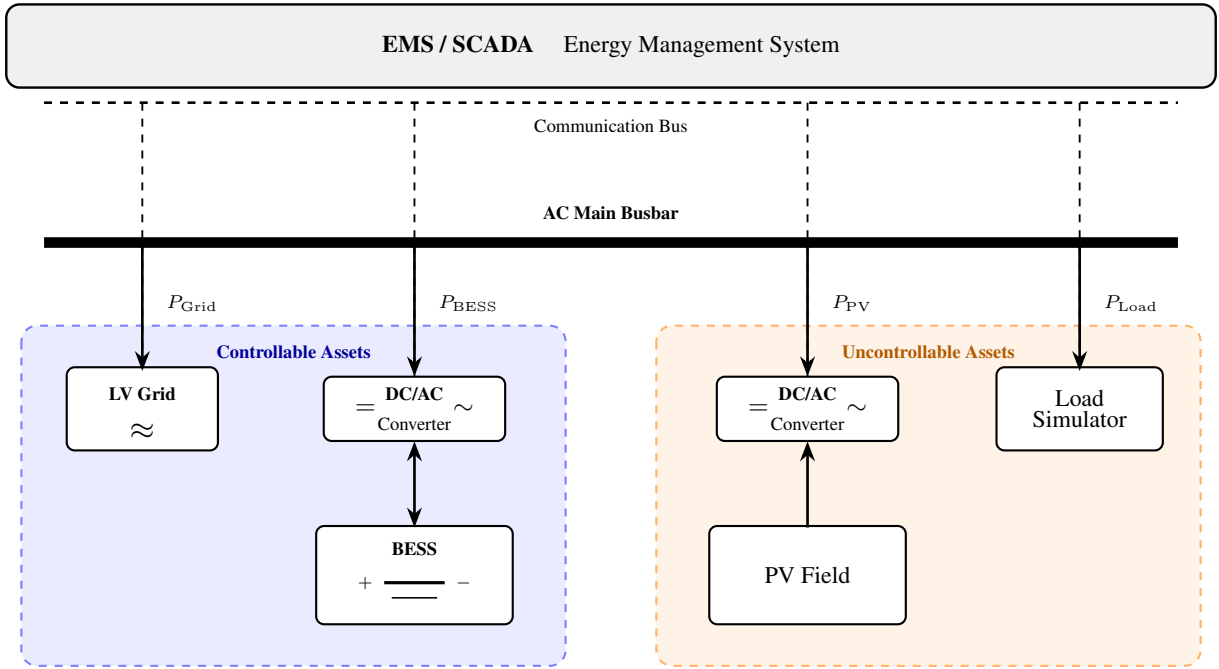
\begin{figure*}[!ht]
\centering
\begin{tikzpicture}[
    font=\small,
    >=Stealth,
    every node/.style={align=center},
    box/.style={draw, rounded corners=3pt, minimum width=2.4cm, minimum height=1.0cm, fill=white, line width=0.8pt},
    converter/.style={draw, rounded corners=3pt, minimum width=2.4cm, minimum height=0.85cm, fill=white, line width=0.8pt},
]

\begin{scope}[on background layer]
    \fill[blue!8, rounded corners=6pt]
        (-7.8, -1.1) rectangle (-0.6, -5.6);
    \fill[orange!10, rounded corners=6pt]
        (0.6, -1.1) rectangle (7.8, -5.6);
\end{scope}

\draw[blue!50, dashed, rounded corners=6pt, line width=0.8pt]
    (-7.8, -1.1) rectangle (-0.6, -5.6);
\draw[orange!60, dashed, rounded corners=6pt, line width=0.8pt]
    (0.6, -1.1) rectangle (7.8, -5.6);

\node[font=\scriptsize\bfseries, text=blue!60!black] at (-4.2, -1.45) {Controllable Assets};
\node[font=\scriptsize\bfseries, text=orange!70!black] at (4.2, -1.45) {Uncontrollable Assets};

\draw[line width=4pt] (-7.5, 0) -- (7.5, 0);
\node[font=\scriptsize\bfseries] at (0, 0.4) {AC Main Busbar};

\node[box, minimum width=2.0cm, minimum height=1.1cm] (grid) at (-6.2, -2.2) {};
\node[font=\scriptsize\bfseries] at (-6.2, -2.0) {LV Grid};
\node[font=\large] at (-6.2, -2.5) {$\approx$};

\draw[->, line width=1.0pt] (-6.2, -0.05) -- (-6.2, -1.7);
\node[font=\scriptsize, right] at (-6.0, -0.8) {$P_{\mathrm{Grid}}$};

\node[converter] (conv_bess) at (-2.6, -2.2) {};
\node[font=\scriptsize\bfseries] at (-2.6, -2.0) {DC/AC};
\node[font=\scriptsize] at (-2.6, -2.4) {Converter};
\node[font=\normalsize] at (-3.25, -2.2) {$=$};
\node[font=\normalsize] at (-1.95, -2.2) {$\sim$};

\node[box, minimum width=2.6cm, minimum height=1.3cm] (bess) at (-2.6, -4.4) {};
\node[font=\scriptsize\bfseries] at (-2.6, -4.05) {BESS};
\draw[line width=1.2pt] (-3.0,-4.5) -- (-2.2,-4.5);
\draw[line width=0.6pt] (-2.9,-4.7) -- (-2.3,-4.7);
\node[font=\tiny] at (-3.25, -4.5) {$+$};
\node[font=\tiny] at (-1.95, -4.5) {$-$};

\draw[->, line width=1.0pt] (-2.6, -0.05) -- (-2.6, -1.77);
\node[font=\scriptsize, right] at (-2.4, -0.8) {$P_{\mathrm{BESS}}$};

\draw[<->, line width=1.0pt] (-2.6, -2.63) -- (-2.6, -3.77);

\node[converter] (conv_pv) at (2.6, -2.2) {};
\node[font=\scriptsize\bfseries] at (2.6, -2.0) {DC/AC};
\node[font=\scriptsize] at (2.6, -2.4) {Converter};
\node[font=\normalsize] at (1.95, -2.2) {$=$};
\node[font=\normalsize] at (3.25, -2.2) {$\sim$};

\node[box, minimum width=2.6cm, minimum height=1.3cm] (pv) at (2.6, -4.4) {PV Field};

\draw[->, line width=1.0pt] (2.6, -0.05) -- (2.6, -1.77);
\node[font=\scriptsize, right] at (2.8, -0.8) {$P_{\mathrm{PV}}$};

\draw[->, line width=1.0pt] (2.6, -3.77) -- (2.6, -2.63);

\node[box, minimum width=2.2cm, minimum height=1.1cm] (load) at (6.2, -2.2) {Load\\[-2pt]Simulator};
\draw[->, line width=1.0pt] (6.2, -0.05) -- (6.2, -1.7);
\node[font=\scriptsize, right] at (6.4, -0.8) {$P_{\mathrm{Load}}$};

\node[draw, rounded corners=5pt, fill=gray!12, line width=0.9pt,
      minimum width=16.0cm, minimum height=1.1cm] (ems) at (0, 2.6)
      {\textbf{EMS / SCADA} \quad Energy Management System};

\draw[dashed, line width=1.0pt] (-7.5, 1.85) -- (7.5, 1.85);
\node[font=\scriptsize] at (0, 1.55) {Communication Bus};

\draw[dashed, line width=0.7pt] (-6.2, 1.85) -- (-6.2, -1.7);
\draw[dashed, line width=0.7pt] (-2.6, 1.85) -- (-2.6, -1.77);
\draw[dashed, line width=0.7pt]  (2.6, 1.85) -- (2.6,  -1.77);
\draw[dashed, line width=0.7pt]  (6.2, 1.85) -- (6.2,  -1.7);

\end{tikzpicture}
\caption{MG$^2$Lab microgrid architecture. The BESS and grid connection constitute the controllable assets of the EMS, while PV generation and load demand are treated as uncontrollable uncertainty sources.}
\label{fig:microgrid}
\end{figure*}

The microgrid is connected to the main distribution grid through a bidirectional grid connection, with maximum import power $P_{\text{buy},\text{max}}$ and maximum export power $P_{\text{sell},\text{max}}$. The layout of the MG is illustrated in the Fig. \ref{fig:microgrid}.

\subsection{Energy Management System Formulation}
\label{text3:ems}

The microgrid EMS is formulated as a two-stage linear program (LP) that mirrors real electricity market operations. In the day-ahead (DA) stage, the BESS charge and discharge schedule is committed from the forecast inputs $\hat{p}_{\text{pv},t}$ and~$\hat{y}_t$ for PV generation and load demand, respectively. In the real-time (RT) stage, the committed schedule is fixed and the system re-dispatches the available degrees of freedom, namely grid exchange and load shedding, against the actual realizations $p_{\text{pv},t}$ and $y_t$. This sequential structure ensures that forecast errors propagate into real operational consequences, providing an economically meaningful signal for model training.

\subsubsection{Day-Ahead Optimization}
\label{text3:DA}

The DA stage takes the forecast pair $(\hat{p}_{\text{pv},t}, \hat{y}_t)$ as input and seeks to minimize the following objective over the scheduling horizon $\mathcal{T}$:
\begin{equation}
\begin{split}
\min_{\substack{\mathbf{P}_{\text{buy}}, \mathbf{P}_{\text{sell}},
    \mathbf{P}_{\text{ch}}, \mathbf{P}_{\text{dis}}, \\
    \mathbf{SOE}, \mathbf{P}_{\text{curt}}}}
\sum_{t \in \mathcal{T}} \Big[
    &\lambda_{\text{buy},t} P_{\text{buy},t} \Delta t
    - \lambda_{\text{sell},t} P_{\text{sell},t} \Delta t \\
    &+ c_{\text{OM}} P_{\text{dis},t} \Delta t
    + c_{\text{curt}} P_{\text{curt},t} \Delta t
\Big],
\end{split}
\label{text3:eq:da_obj}
\end{equation}
where $P_{\text{buy},t}$ and $P_{\text{sell},t}$ are the power purchased from and sold to the grid in period $t$; $\mathbf{P}_{\text{ch}}$ and $\mathbf{P}_{\text{dis}}$ are the BESS charging and discharging power profiles; $\mathbf{SOE}$ is the state of energy of the BESS; $\mathbf{P}_{\text{curt}}$ is the curtailed PV power, with $P_{\text{dis},t}$ and $P_{\text{curt},t}$ their values in period $t$; $\lambda_{\text{buy},t}$ and $\lambda_{\text{sell},t}$ are the time-varying grid purchase and sell prices; $c_{\text{OM}}$ is the BESS operation and maintenance cost per unit of discharged energy; $c_{\text{curt}}$ is the penalty per unit of curtailed PV energy; and $\Delta t$ is the duration of each time step. The objective function~\eqref{text3:eq:da_obj} minimizes the sum of grid purchase costs, net of grid sell revenues, plus BESS operation and maintenance costs and PV curtailment penalties.

The DA stage is subject to the following constraints. The power balance enforces that, at every timestep, total supply equals total demand:
\begin{equation}
\begin{split}
\hat{p}_{\text{pv},t} - P_{\text{curt},t} + \eta_{\text{dis}}\, P_{\text{dis},t}
   - \frac{P_{\text{ch},t}}{\eta_{\text{ch}}} \\
{}+ P_{\text{buy},t} - P_{\text{sell},t} &= \hat{y}_t,
   \quad \forall t \in \mathcal{T}, 
\end{split}
\label{text3:eq:da_balance}
\end{equation}
where $\hat{p}_{\text{pv},t}$ is the day-ahead PV generation forecast, $\hat{y}_t$ is the day-ahead load forecast, and $\eta_{\text{ch}}$, $\eta_{\text{dis}}$ are the BESS charge and discharge efficiencies, respectively.

The BESS SOE evolves according to its charge and discharge schedule and is restricted to its usable energy window:
\begin{subequations}\label{text3:eq:soe}
\begin{align}
SOE_t &=
\begin{cases}
SOE_{\text{init}} + \left(P_{\text{ch},0} - P_{\text{dis},0}\right)\Delta t, & t = 0, \\[4pt]
SOE_{t-1} + \left(P_{\text{ch},t} - P_{\text{dis},t}\right)\Delta t, & t > 0,
\end{cases}
\label{text3:eq:soe_dyn} \\[4pt]
SOE_{\text{min}} &\leq SOE_t \leq SOE_{\text{max}}, \qquad \forall t \in \mathcal{T}.
\label{text3:eq:soe_bounds}
\end{align}
\end{subequations}
The remaining decision variables are constrained by the physical limits of the BESS, the grid connection, and the available PV generation:
\begin{subequations}\label{text3:eq:da_bounds}
\begin{align}
0 &\leq P_{\text{ch},t} \leq P_{\text{ch},\text{max}}, \qquad \forall t \in \mathcal{T}, \label{text3:eq:da_ch} \\
0 &\leq P_{\text{dis},t} \leq P_{\text{dis},\text{max}}, \qquad \forall t \in \mathcal{T}, \label{text3:eq:da_dis} \\
0 &\leq P_{\text{buy},t} \leq P_{\text{buy},\text{max}}, \qquad \forall t \in \mathcal{T}, \label{text3:eq:da_buy} \\
0 &\leq P_{\text{sell},t} \leq P_{\text{sell},\text{max}}, \qquad \forall t \in \mathcal{T}, \label{text3:eq:da_sell} \\
0 &\leq P_{\text{curt},t} \leq \hat{p}_{\text{pv},t}, \qquad \forall t \in \mathcal{T}. \label{text3:eq:da_curt}
\end{align}
\end{subequations}
The solution to the DA problem yields the committed BESS charge and discharge schedules $P_{\text{ch},t}^{*}$ and $P_{\text{dis},t}^{*}$ for all $t \in \mathcal{T}$. These schedules are fixed prior to real-time operation and cannot be revised once the actual load and PV generation are revealed.

\subsubsection{Real-Time Optimization}
\label{text3:RT}

Once the committed BESS schedule $\left( P_{\text{ch}}^{*}, P_{\text{dis}}^{*} \right)$ is set and the actual load $y_t$ and actual PV generation $p_{\text{pv},t}$ are observed, the RT stage re-optimizes the remaining controllable variables to minimize real operational costs derived as:
\begin{equation}
\begin{split}
\min_{\substack{\mathbf{P}_{\text{buy}}, \mathbf{P}_{\text{sell}}, \\
    \mathbf{P}_{\text{curt}}, \mathbf{P}_{\text{shed}}}}
\sum_{t \in \mathcal{T}} \Big[
    &\lambda_{\text{buy},t} P_{\text{buy},t} \Delta t
    - \lambda_{\text{sell},t} P_{\text{sell},t} \Delta t \\
    &+ c_{\text{OM}} P_{\text{dis},t}^{*} \Delta t
    + c_{\text{curt}} P_{\text{curt},t} \Delta t \\
    &+ c_{\text{shed}} P_{\text{shed},t} \Delta t
\Big]
\end{split}
\label{text3:eq:rt_obj}
\end{equation}
The BESS schedule enters the RT objective~\eqref{text3:eq:rt_obj} as a fixed parameter through the O\&M cost term $c_{\text{OM}} P_{\text{dis},t}^{*}$, reflecting the wear cost incurred regardless of forecast quality. The power balance is modified to incorporate the actual realizations of both uncertain sources and the load shedding variable $P_{\text{shed},t}$:
\begin{equation}
\begin{split}
&p_{\text{pv},t} - P_{\text{curt},t} + \eta_{\text{dis}}\, P_{\text{dis},t}^{*}
   - \frac{P_{\text{ch},t}^{*}}{\eta_{\text{ch}}} \\
&\quad + P_{\text{buy},t} - P_{\text{sell},t} + P_{\text{shed},t} = y_t,
   \quad \forall t \in \mathcal{T}
\end{split}
\label{text3:eq:rt_balance}
\end{equation}
Comparing~\eqref{text3:eq:da_balance} and~\eqref{text3:eq:rt_balance}, two structural differences emerge. First, the forecast inputs $\hat{p}_{\text{pv},t}$ and $\hat{y}_t$ are replaced by their actual realizations $p_{\text{pv},t}$ and $y_t$, respectively. Second, the BESS charge and discharge terms become fixed parameters $P_{\text{ch},t}^{*}$ and $P_{\text{dis},t}^{*}$, and the load shedding slack variable $P_{\text{shed},t}$ is introduced to guarantee feasibility. The remaining constraints enforce non-negativity and physical limits on grid exchange, curtailment, and load shedding:
\begin{subequations}\label{text3:eq:rt_bounds}
\begin{align}
0 \leq {}&P_{\text{buy},t} \leq P_{\text{buy},\text{max}},   &&\forall t \in \mathcal{T}, \label{text3:eq:rt_buy} \\
0 \leq {}&P_{\text{sell},t} \leq P_{\text{sell},\text{max}}, &&\forall t \in \mathcal{T}, \label{text3:eq:rt_sell} \\
0 \leq {}&P_{\text{curt},t} \leq p_{\text{pv},t},            &&\forall t \in \mathcal{T}, \label{text3:eq:rt_curt} \\
&P_{\text{shed},t} \geq 0,                                   &&\forall t \in \mathcal{T}. \label{text3:eq:rt_shed}
\end{align}
\end{subequations}
Note that the curtailment bound~\eqref{text3:eq:rt_curt} now references the actual PV generation $p_{\text{pv},t}$ rather than the forecast $\hat{p}_{\text{pv},t}$ used in the DA stage~\eqref{text3:eq:da_curt}. This distinction is operationally significant: at the DA stage the system can only plan curtailment against expected generation, whereas at real time it curtails against what is physically available.


\section{Decision-Focused Learning Framework}
\label{text4:DFL}

The proposed DFL framework integrates forecasting and MG optimization into a single E2E trainable pipeline. Rather than minimizing a generic statistical loss such as the mean squared error (MSE), the forecasting model is trained directly on the real operational cost produced by the two-stage EMS described in Section~\ref{text3:ems}. This is achieved by embedding the DA and RT optimization problems as differentiable layers within the neural network training graph, allowing gradients to flow from the real-time operational cost back through the optimization and into the forecaster weights.

The framework consists of three components: a forecasting model, a differentiable optimization layer, and a cost-based loss function. The forecasting model comprises two independent gated recurrent unit (GRU) networks that produce the day-ahead load forecast $\hat{\mathbf{y}} \in \mathbb{R}^{T}$ and the PV generation forecast $\hat{\mathbf{p}}_{\text{pv}} \in \mathbb{R}^{T}$ from a window of past measurements augmented with cyclical hour-of-day and day-of-week encodings, with a ReLU output layer \cite{ReLu} enforcing the physical non-negativity of both quantities. The two forecasters share the same architecture and are trained independently. The remaining two components are described in the following subsections, with the forecaster architecture and all hyperparameters deferred to Section~\ref{subsec:models}.

\begin{figure*}
	\centering
	\includegraphics[width=\textwidth]{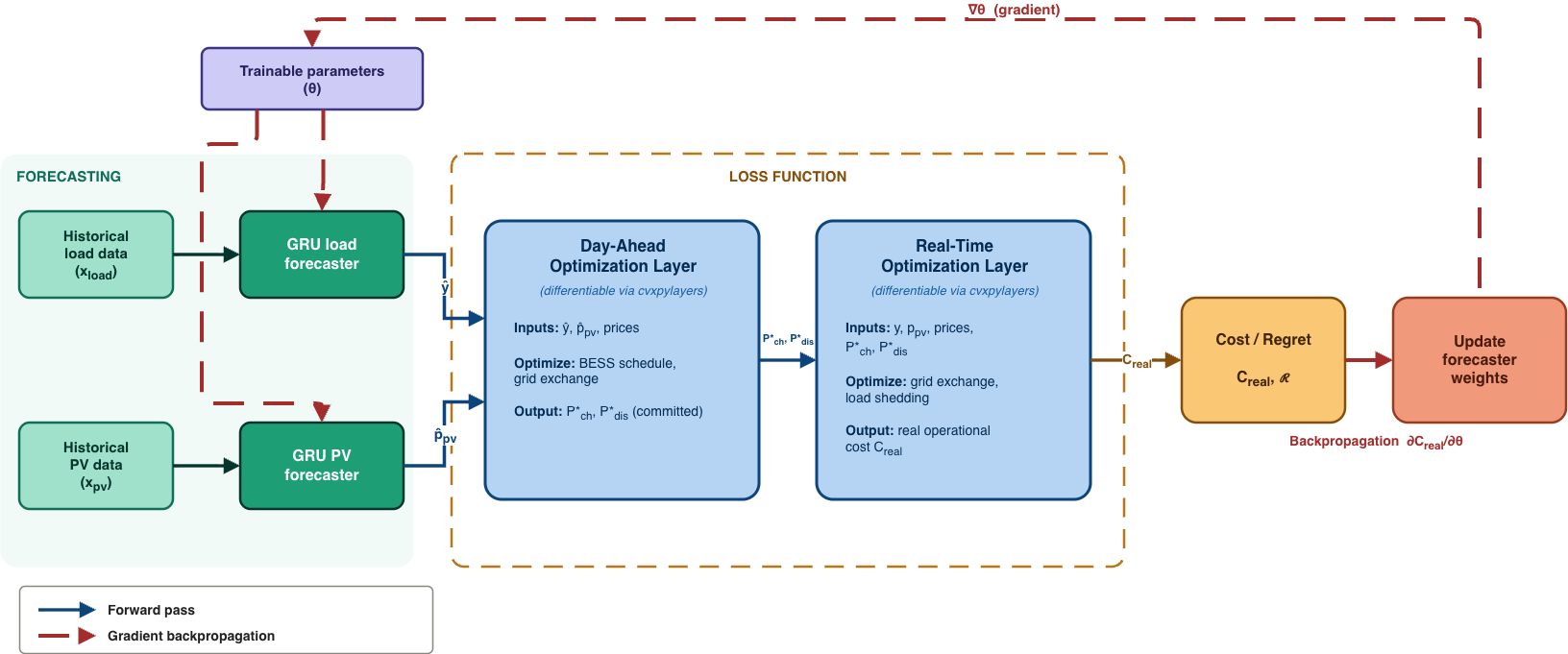}
	\caption{Flow graph demonstrating the end-to-end learning architecture of the forecast modules.}
	\label{text4:fig:flowgraph}
\end{figure*}

\subsection{Differentiable Optimization Layer}
\label{text4:DOL}

A fundamental challenge in DFL is that optimization problems are not natively differentiable: standard solvers produce optimal solutions $x^*$ but provide no gradient information with respect to the input parameters. To train the GRU forecasters end-to-end with the EMS, it is necessary to backpropagate gradients through the optimization itself, computing how changes in the forecasts affect the committed BESS schedule and, consequently, the real operational cost.

This is achieved through the \texttt{cvxpylayers} library \cite{cvxpylayers2019}, which embeds convex optimization problems as differentiable layers within a \texttt{PyTorch} computational graph. Both the DA and RT optimization problems described in Section~\ref{text3:ems} are implemented as \texttt{cvxpylayers} instances. Internally, \texttt{cvxpylayers} computes exact gradients of the optimal solution with respect to the problem parameters via implicit differentiation of the Karush-Kuhn-Tucker (KKT) optimality conditions \cite{cvxpylayers2019}, solving a linear system at each backward pass to provide exact sensitivities without finite differences or surrogate approximations. The splitting conic solver \texttt{(SCS)} is used as the underlying solver.

During the forward pass, \texttt{SCS} solves the DA problem given both the load forecast $\hat{\mathbf{y}}$ and the PV forecast $\hat{\mathbf{p}}_{\text{pv}}$, and returns the optimal BESS commitment $\left(P_{\text{ch}}^*, P_{\text{dis}}^*\right)$. The committed schedule is then passed as a fixed parameter to the RT layer, which solves the real-time problem against the actual load $\mathbf{y}$ and actual PV generation $\mathbf{p}_{\text{pv}}$ and returns the operational cost $C_{\text{real}}$. Because the two forecasters are trained independently, with the complementary input held at its realized value, we denote by $\hat{\mathbf{z}} \in \{\hat{\mathbf{y}}, \hat{\mathbf{p}}_{\text{pv}}\}$ the forecast of the variable currently being trained. The backward pass then produces the gradient chain:
\begin{equation}
\frac{\partial C_{\text{real}}}{\partial \theta} = 
\frac{\partial C_{\text{real}}}{\partial P_{\text{ch}}^*}
\frac{\partial P_{\text{ch}}^*}{\partial \hat{\mathbf{z}}}
\frac{\partial \hat{\mathbf{z}}}{\partial \theta} + 
\frac{\partial C_{\text{real}}}{\partial P_{\text{dis}}^*}
\frac{\partial P_{\text{dis}}^*}{\partial \hat{\mathbf{z}}}
\frac{\partial \hat{\mathbf{z}}}{\partial \theta}
\label{text4:eq:chain}
\end{equation}
where $\partial P_{\text{ch}}^*/\partial \hat{\mathbf{z}}$ and $\partial P_{\text{dis}}^*/\partial \hat{\mathbf{z}}$ are computed by \texttt{cvxpylayers} via KKT-based implicit differentiation, and $\partial \hat{\mathbf{z}}/\partial \theta$ by \texttt{PyTorch} autograd through the GRU. The load forecast enters the DA problem through the power balance~\eqref{text3:eq:da_balance}, and the PV forecast through both the power balance and the curtailment bound~\eqref{text3:eq:da_curt}; in either case the forecast parameterizes the committed schedule, so the same chain applies with $\hat{\mathbf{z}} = \hat{\mathbf{y}}$ to train the load forecaster and $\hat{\mathbf{z}} = \hat{\mathbf{p}}_{\text{pv}}$ to train the PV forecaster, as illustrated in Fig. \ref{text4:fig:flowgraph}. This gradient path allows each forecaster to learn which prediction errors are operationally costly and adjust its weights accordingly.

Note that \texttt{cvxpylayers} requires the optimization problem to be convex. Both the DA and RT formulations in Section~\ref{text3:ems} satisfy this requirement, as all objective and constraint functions are linear in the decision variables.

\subsection{Loss Function and Training Procedure}
\label{text4:loss}

The loss function adopted for DFL training is the real operational cost $C_{\text{real}}$ produced by the RT optimization layer. For the forecaster under training, the objective is:
\begin{equation}
\min_{\theta} \quad \mathbb{E}_{(\mathbf{x},\mathbf{z}) \sim \mathcal{D}} 
\left[ C_{\text{real}}\left(\hat{\mathbf{z}}(\mathbf{x};\theta), \mathbf{z}\right) 
\right],
\label{text4:eq:dfl_obj}
\end{equation}
where $\theta$ denotes the GRU parameters, $x$ the input feature sequence, $z$ the realization of the forecast variable (load or PV), $\mathcal{D}$ denotes the underlying data distribution and $\hat{\mathbf{z}}(x;\theta)$ its forecast, with the complementary input held fixed. To quantify the suboptimality introduced by forecast errors, the regret is defined as the additional cost incurred relative to a perfect foresight benchmark:
\begin{equation}
\mathcal{R}(\theta) = C_{\text{real}}
\left(\hat{\mathbf{z}}(x;\theta), z\right) - 
C_{\text{pf}}(z), 
\label{text4:eq:regret}
\end{equation}
where $C_{\text{pf}}(z)$ is the perfect foresight cost obtained by solving the DA stage with the realization $z$ in place of the forecast. Because $C_{\text{pf}}$ does not depend on $\theta$, minimizing \eqref{text4:eq:dfl_obj} and \eqref{text4:eq:regret} is equivalent. The regret is used exclusively as an evaluation metric to assess the gap between the trained model and the theoretical optimum.
The structure of $C_{\text{real}}$ creates an inherently asymmetric training signal, so the forecaster learns to bias its predictions in the direction that minimizes shedding risk, producing forecasts that are statistically suboptimal yet operationally superior.
A known challenge in DFL training is the cold-start problem: with randomly initialized weights, the forecaster produces poor predictions that may generate large unstable gradients, destabilizing training from the outset \cite{DFL_KULeuven}. To address this, a two-phase strategy is adopted. In the first phase, the forecaster is pretrained for $E_{\text{warmup}}$ epochs using the standard MSE loss:
\begin{equation}
\mathcal{L}_{\text{MSE}}(\theta) = \frac{1}{T} 
\sum_{t=0}^{T-1} \left\| \hat{z}_t(\theta) - z_t \right\|_2^2,
\label{text4:eq:mse}
\end{equation}
providing a reasonable starting point before DFL training begins. In the second phase, the MSE loss is discarded and the forecaster is trained end-to-end by minimizing $C_{\text{real}}$ as in \eqref{text4:eq:dfl_obj}, with gradients flowing through both optimization layers back to the GRU weights via~\eqref{text4:eq:chain}. The complete procedure is summarized in Algorithm~\ref{text4:alg:dfl_full}.

\begin{algorithm}[ht]
\DontPrintSemicolon
\SetAlgoLined
\small
\KwIn{$\mathcal{D} = \{(\mathbf{x}_i, \mathbf{z}_i, \mathbf{w}_i, \boldsymbol{\lambda}_i)\}_{i=1}^{N}$, with $\mathbf{z}$ the target realization, $\mathbf{w}$ the fixed complementary input, and $\boldsymbol{\lambda}$ the vector of grid prices (purchase and sell); $E_{\text{warmup}}$, $E_{\text{max}}$, $\alpha$, $P$,}
\KwOut{$\theta^*$ and test metrics $C_{\text{real}}, \mathcal{R}, P_{\text{shed}}$}
\BlankLine
\tcp{Phase 1: MSE warmup}
\For{$e = 1$ \KwTo $E_{\text{warmup}}$}{
    \For{$(\mathbf{x}, \mathbf{z}) \in \mathcal{D}$}{
        $\hat{\mathbf{z}} \leftarrow \text{ReLU}(\text{GRU}(\mathbf{x};\theta))$\;
        $\theta \leftarrow \theta - \alpha \nabla_\theta \mathcal{L}_{MSE}$ \tcp*{Eq.~\eqref{text4:eq:mse}}
    }
}
\BlankLine
\tcp{Phase 2: DFL training}
$p \leftarrow 0$\;
\For{$e = 1$ \KwTo $E_{\text{max}}$}{
    \For{$(\mathbf{x}, \mathbf{z}, \mathbf{w}, \boldsymbol{\lambda}) \in \mathcal{D}$}{
        $\hat{\mathbf{z}} \leftarrow \text{ReLU}(\text{GRU}(\mathbf{x};\theta))$\;
        $P_{\text{ch}}^*, P_{\text{dis}}^* \leftarrow \text{DA-Layer}(\hat{\mathbf{z}}, \mathbf{w}, \boldsymbol{\lambda})$ \tcp*{\eqref{text3:eq:da_balance}--\eqref{text3:eq:da_curt}}
        $C_{\text{real}}, P_{\text{shed}} \leftarrow \text{RT-Layer}(\mathbf{z}, \mathbf{w}, \boldsymbol{\lambda}, P_{\text{ch}}^*, P_{\text{dis}}^*)$ \tcp*{\eqref{text3:eq:rt_balance}--\eqref{text3:eq:rt_shed}}
        $\theta \leftarrow \theta - \alpha \, \text{clip}\!\left(\nabla_\theta C_{\text{real}}\right)$ \tcp*{via~\eqref{text4:eq:chain}}
    }
    \tcp{Early stopping on validation $C_{\text{real}}$}
    \eIf{val. $C_{\text{real}}$ improved}{
        $p \leftarrow 0$, \ $\theta^* \leftarrow \theta$\;
    }{
        $p \leftarrow p + 1$\; \lIf{$p \geq P$}{\textbf{break}}
    }
}
\BlankLine
\tcp{Phase 3: deployment and evaluation}
\For{$(\mathbf{x}, \mathbf{z}, \mathbf{w}, \boldsymbol{\lambda})$ in test set}{
    $\hat{\mathbf{z}} \leftarrow \text{ReLU}(\text{GRU}(\mathbf{x};\theta^*))$\;
    $P^* \leftarrow \text{DA-Layer}(\hat{\mathbf{z}}, \mathbf{w}, \boldsymbol{\lambda})$;\quad $C_{\text{real}}, P_{\text{shed}} \leftarrow \text{RT-Layer}(\mathbf{z}, \mathbf{w}, \boldsymbol{\lambda}, P^*)$\;
    $P_{\text{pf}}^* \leftarrow \text{DA-Layer}(\mathbf{z}, \mathbf{w}, \boldsymbol{\lambda})$;\quad $C_{\text{pf}} \leftarrow \text{RT-Layer}(\mathbf{z}, \mathbf{w}, \boldsymbol{\lambda}, P_{\text{pf}}^*)$\;
    $\mathcal{R} \leftarrow C_{\text{real}} - C_{\text{pf}}$\;
}
\Return $\theta^*$\;
\caption{End-to-end training and deployment (per forecaster)}
\label{text4:alg:dfl_full}
\end{algorithm}
\section{Experimental Setup and Case Study}
\label{sec:casestudies}

The proposed DFL framework is evaluated on the MG\textsuperscript{2}Lab  
experimental microgrid under two operational scenarios 
of increasing grid constraint: normal operation, and 
grid-constrained operation. 
This progressive structure allows a systematic 
assessment of how the DFL advantage evolves as grid 
flexibility decreases and the BESS commitment decision 
becomes increasingly critical. All case studies share 
the same dataset, model architecture, and training 
procedure described in the current section.

\subsection{System and Model Parameters}
\label{subsec:models}

The physical parameters of the MG$^2$Lab microgrid are summarized in Table~\ref{tab:microgrid_params}. The PV system has a rated capacity of 23 kWp, while the EV charging point imposes a maximum demand of 25 kW. The BESS operates within a SOE range of 25--60 kWh, with symmetric charge and discharge power limits of 40 kW and an efficiency of 0.97 in both directions. An O\&M cost of 0.04~\texteuro \,/kWh is assigned to battery cycling, and curtailment and load shedding penalties are both fixed at 4.0~\texteuro \,/kWh to discourage constraint violations \cite{VoLL_elsevier}. The grid import and export limits differ between the two scenarios and are specified in Section~\ref{sec:scenarios}.

\begin{table}[!ht]
\caption{MG$^2$Lab microgrid system parameters.}
\label{tab:microgrid_params}
\centering
\small
\setlength{\tabcolsep}{4pt}
\begin{tabular}{@{}llrl@{}}
\toprule
\textbf{Parameter} & \textbf{Symbol} & \textbf{Value} & \textbf{Unit} \\
\midrule
\multicolumn{4}{@{}l}{\textit{PV Generation}} \\
\quad Rated capacity & $P_{\text{pv},\max}$ & 23.5 & kWp \\[3pt]
\multicolumn{4}{@{}l}{\textit{Electric Vehicle Charging Point}} \\
\quad Maximum demand & $P_{\text{ev},\max}$ & 25 & kW \\[3pt]
\multicolumn{4}{@{}l}{\textit{Battery Energy Storage System}} \\
\quad Maximum SOE & $SOE_{\max}$ & 60 & kWh \\
\quad Minimum SOE & $SOE_{\min}$ & 25 & kWh \\
\quad Initial SOE & $SOE_{\text{init}}$ & 36 & kWh \\
\quad Maximum charge power & $P_{\text{ch},\max}$ & 40 & kW \\
\quad Maximum discharge power & $P_{\text{dis},\max}$ & 40 & kW \\
\quad Charge efficiency & $\eta_{\text{ch}}$ & 0.97 & -- \\
\quad Discharge efficiency & $\eta_{\text{dis}}$ & 0.97 & -- \\
\quad O\&M cost & $c_{\text{OM}}$ & 0.04 & \euro/kWh \\[3pt]
\multicolumn{4}{@{}l}{\textit{Cost Parameters}} \\
\quad Curtailment penalty & $c_{\text{curt}}$ & 4.0 & \euro/kWh \\
\quad Load shedding penalty & $c_{\text{shed}}$ & 4.0 & \euro/kWh \\[3pt]
\multicolumn{4}{@{}l}{\textit{Grid Connection}} \\
\quad Maximum import & $P_{\text{buy},\max}$ & \multicolumn{2}{l}{see Section~\ref{sec:scenarios}} \\
\quad Maximum export & $P_{\text{sell},\max}$ & \multicolumn{2}{l}{see Section~\ref{sec:scenarios}} \\
\bottomrule
\end{tabular}
\end{table}

The GRU forecaster architecture and training hyperparameters shared across all configurations are reported in Table~\ref{tab:gru_hyperparams}. Each forecaster takes as input a lookback window of three days (288 timesteps at a 15-minute resolution) and produces a full-horizon forecast of 96 timesteps, corresponding to a 24-hour ahead schedule. The network consists of two stacked GRU layers with a hidden dimension of 64 and a dropout rate of 0.2 to mitigate overfitting. Training is performed using the Adam optimizer with a learning rate of $10^{-3}$ and a batch size of 32, for a maximum of 100 epochs with early stopping patience of 10 and gradient clipping set to 1.0 to ensure stable convergence.

\begin{table}[!ht]
\caption{GRU forecaster architecture and training hyperparameters.}
\label{tab:gru_hyperparams}
\centering
\small
\setlength{\tabcolsep}{4pt}
\begin{tabular}{@{}lll@{}}
\toprule
\textbf{Hyperparameter} & \textbf{Symbol} & \textbf{Value} \\
\midrule
\multicolumn{3}{@{}l}{\textit{Architecture}} \\
\quad Lookback window & $L$ & 3 days (288 timesteps) \\
\quad Forecast horizon & $T$ & 96 timesteps \\
\quad Hidden dimension & $d_{h}$ & 64 \\
\quad Number of layers & -- & 2 \\
\quad Dropout rate & -- & 0.2 \\[3pt]
\multicolumn{3}{@{}l}{\textit{Training}} \\
\quad Optimizer & -- & Adam \\
\quad Learning rate & $\alpha$ & $10^{-3}$ \\
\quad Batch size & -- & 32 \\
\quad Maximum epochs & -- & 100 \\
\quad Early stopping patience & -- & 10 \\
\quad Gradient clipping norm & -- & 1.0 \\
\bottomrule
\end{tabular}
\end{table}

\subsection{MG\textsuperscript{2}Lab  Dataset}
\label{sec:data}
The dataset used in this study is collected from the 
MG\textsuperscript{2}Lab  experimental microgrid facility at Politecnico di Milano \cite{MG2Lab}. The dataset comprises three time series recorded at 15-minute resolution: aggregate load demand, PV generation, and electricity prices. The load profile includes the contribution of the EV charging point, which introduces significant stochastic variability and makes day-ahead load forecasting a particularly challenging task.
PV realization is obtained directly from a PV field (23.5 kWp with a 25° tilt and a -6.5° azimuth in 45.50294° N, 9.15658° E) in $\text{MG}^2\text{Lab}$.

The load dataset is the JPL EV charging point, that contains 15-minute intervals of the load power consumption. Specifically, we employ the database created in \cite{impact_mdpi} in turn obtained from \cite{ACN} and scaled to 25 kW. The original database indeed contains the aggregated power demand of 52 charging stations in a workplace parking lot.

\begin{figure*}[ht]
    \centering
    \begin{subfigure}{\textwidth}
        \centering
        \includegraphics[width=1\textwidth]{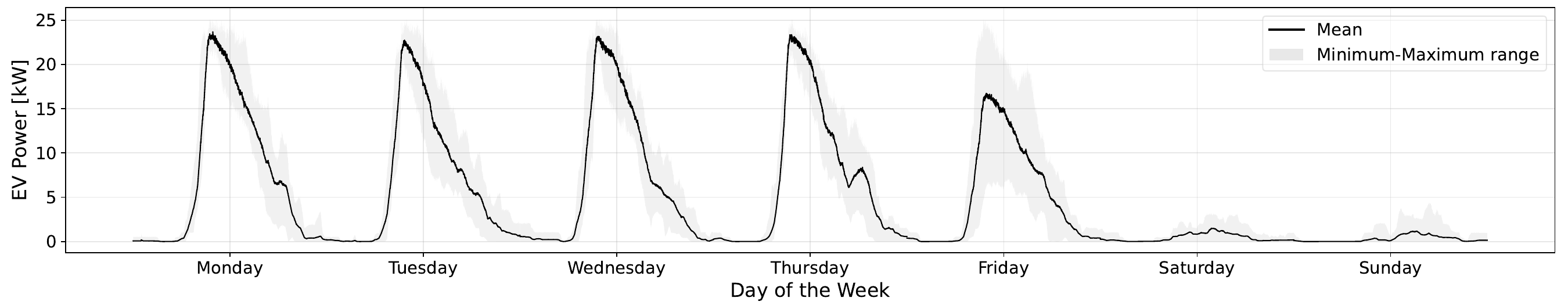}
        \caption{EV charging load profile.}
        \label{fig:ev}
    \end{subfigure}
    \\[1em]
    \begin{subfigure}{0.45\textwidth}
        \centering
        \includegraphics[width=\textwidth, height=5cm]{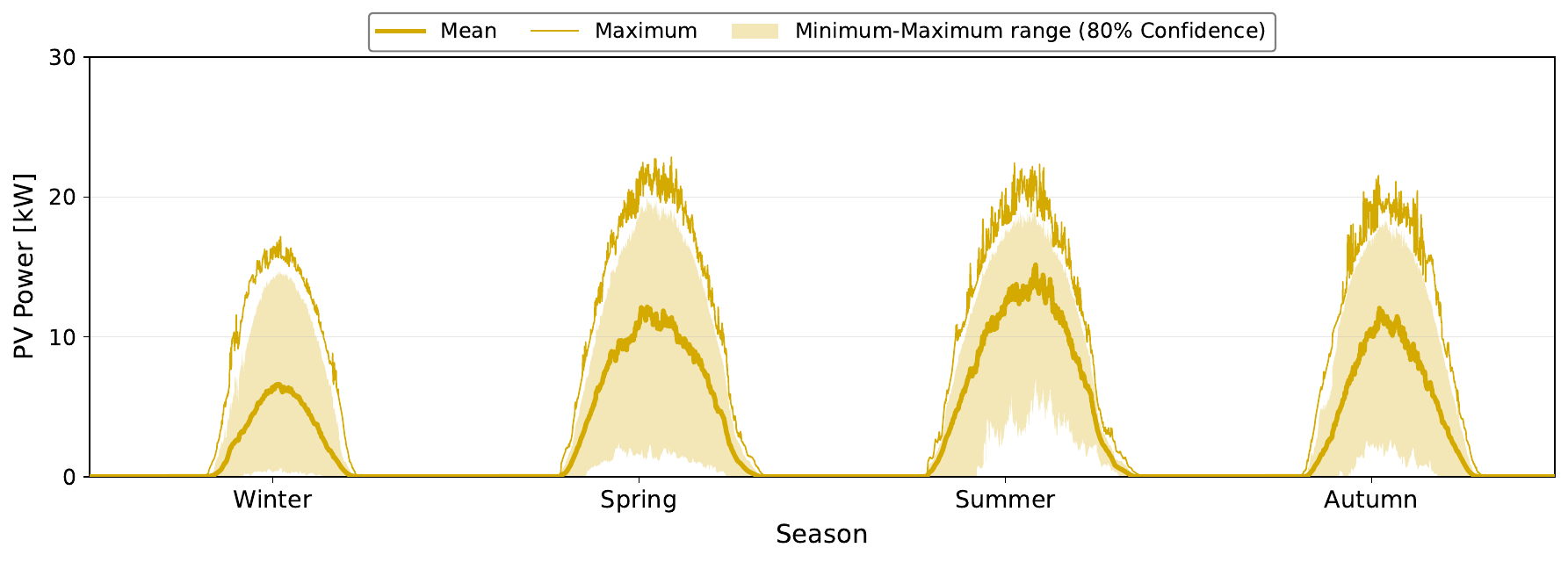}
        \caption{PV generation profile.}
        \label{fig:pv}
    \end{subfigure}
    \hfill
    \begin{subfigure}{0.48\textwidth}
        \centering
        \includegraphics[width=\textwidth]{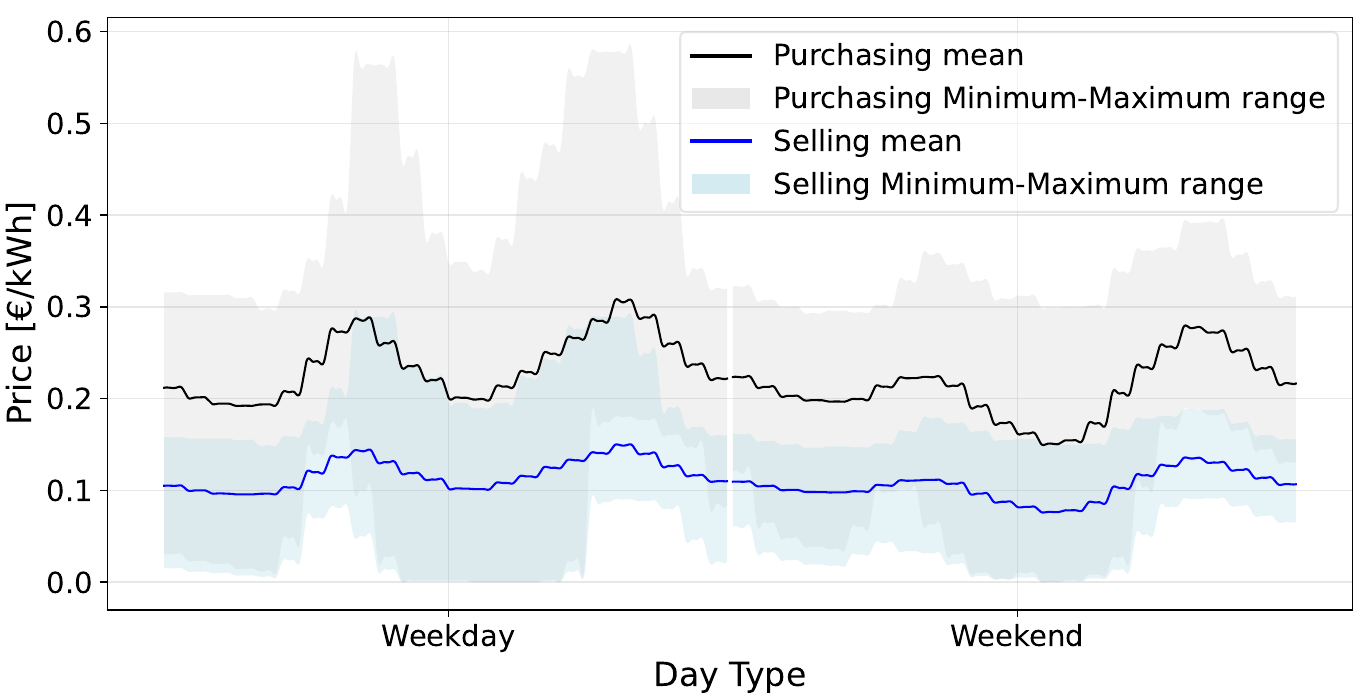}
        \caption{Electricity price profile.}
        \label{fig:prices}
    \end{subfigure}
    \caption{MG\textsuperscript{2}Lab  dataset overview.}
    \label{fig:dataset}
\end{figure*}

In this experiment, prices are considered as deterministic. Therefore, as in \cite{impact_mdpi}, a price database is created where the purchasing price is twice the Italian
national unified price to account for non-energy-related costs in the electricity bill, while the selling price is the
North region zonal price \cite{GME}. We include in Fig. \ref{fig:dataset} a representation of the databases employed in the numerical simulations. Fig. \ref{fig:ev} shows the load power variability and its mean for each day of the week, to highlight the weekly seasonality. Fig. \ref{fig:pv} shows the PV power variability with a 80\% confidence, the maximum registered realization and its mean for each season of the year, to highlight the yearly seasonality. Lastly, Fig. \ref{fig:prices} shows the purchasing and selling price variabilities and their means, differentiating between weekday and weekend.

The total available data spans 464 days, of which 
400 days are used for training and the remaining 
64 days constitute the test set. From the 400 
training days, 20\% are reserved as a validation 
set for early stopping and hyperparameter 
monitoring, leaving 320 days for effective model 
training. The train/test split is chosen to ensure 
that the test period covers a representative range 
of seasonal conditions. Table~\ref{tab:dataset} 
further describes the dataset.

\begin{table}[!ht]
\centering
\caption{MG\textsuperscript{2}Lab  dataset statistics.}
\label{tab:dataset}
\begin{tabular}{@{}lll@{}}
\toprule
\textbf{Property} & \textbf{Value} & \textbf{Unit} \\
\midrule
\multicolumn{3}{@{}l}{\textit{General}} \\
Resolution        & 15  & min  \\
Training days     & 320 & days \\
Validation days   & 80  & days \\
Test days         & 64  & days \\
\midrule
\multicolumn{3}{@{}l}{\textit{Load}} \\
Installed capacity & 25 & kW    \\
Mean demand       & 4.72  & kW   \\
Peak demand       & 25  & kW   \\
\midrule
\multicolumn{3}{@{}l}{\textit{PV Generation}} \\
Installed capacity & 23.5 & kWp   \\
Mean generation   & 2.42  & kW   \\
Peak generation   & 22.72  & kW   \\
\midrule
\multicolumn{3}{@{}l}{\textit{Electricity Prices}} \\
Mean buy price    & 0.229  & \texteuro/kWh \\
Mean sell price   & 0.114  & \texteuro/kWh \\
\bottomrule
\end{tabular}
\end{table}

\subsection{Experimental Scenarios}
\label{sec:scenarios}
Two operational cases of increasing grid 
constraint are defined to systematically evaluate 
the proposed DFL framework. The cases share 
identical dataset, model architecture, training 
procedure, and evaluation protocol, differing 
exclusively in the grid connection parameters 
$P_{\text{buy},\text{max}}$ and $P_{\text{sell},\text{max}}$. This controlled 
design allows a direct assessment of how the DFL 
advantage over accuracy-driven benchmarks evolves 
as grid flexibility decreases and the BESS 
commitment decision becomes progressively more 
critical.

In the first case (Case 1), the MG operates 
under normal grid-connected conditions with import and export limits of 
$P_{\text{buy},\text{max}} = P_{\text{sell},\text{max}} = 40$ kW. The 
grid connection provides sufficient flexibility 
to compensate for suboptimal BESS commitments 
resulting from forecast errors. Load 
shedding events are therefore relatively 
infrequent, and the gradient signal driving 
DFL training is predominantly shaped by cost 
differences rather than shedding penalties. 
This scenario establishes the baseline 
operational performance of the proposed 
framework under standard operating conditions 
and serves as the reference point for the 
subsequent constrained scenarios.

In the second case (Case 2), the grid connection 
capacity is reduced to $P_{\text{buy},\text{max}} = P_{\text{sell},\text{max}} = 18$ kW, reflecting conditions 
of network congestion or a limited grid 
connection agreement. The reduced import 
capacity restricts the ability of the system 
to compensate for poor BESS commitments 
through grid exchange, increasing the 
likelihood and severity of load shedding 
events when the day-ahead load forecast is 
inaccurate. The BESS commitment decision 
therefore carries greater operational 
consequences, and the asymmetric shedding 
penalty $c_{\text{shed}}$ provides a stronger and 
more frequent gradient signal to the DFL 
forecaster relative to Case 1. This case 
is designed to test whether the performance 
advantage of DFL over accuracy-driven 
approaches systematically increases as grid 
flexibility is reduced.

\section{Results and Discussion}
\label{sec:results}

The central question motivating this experimental evaluation is whether training a forecaster to minimize operational cost rather than statistical accuracy translates into measurable economic benefits, and if so, whether this advantage is sensitive to the degree of operational flexibility available to the microgrid. Section 6.1 presents the results under normal operation (40 kW), where the grid provides ample flexibility to absorb suboptimal BESS commitments. Section 6.2 examines the grid-constrained scenario (18 kW), where this flexibility is substantially reduced and forecast errors propagate more directly into load shedding events. Section 6.3 synthesizes both cases through a paired comparison framework to isolate the marginal contribution of decision-focused training on each uncertainty source and quantify the relationship between more stringent constraints and the benefits of a DFL-based approach.

\subsection{Case 1 }
\label{sec:case1}
Under normal operating conditions, the oracle configuration (S9) yields a total operational cost of 1,754.52 \texteuro \, over the 64-day test horizon, as reported in Table \ref{tab:results_40kw}. This value represents the minimum achievable cost when both load demand and PV generation are perfectly known at the day-ahead stage, and constitutes the lower bound against which the regret of all forecasting configurations is evaluated.

To disentangle the individual contributions of load and PV uncertainty, the single-source scenarios (S1–S4) isolate each forecasting source by replacing the other with its actual realization. This controlled design serves two purposes: assessing whether DFL training improves performance on each uncertainty source independently, and quantifying their relative impact on operational cost. The first question is addressed by comparing DFL and MSE configurations within each source. On the PV side, the DFL-trained forecaster reduces cost by 3.01\% relative to its MSE counterpart (S1 vs S3), while on the load side the corresponding improvement is 3.16\% (S2 vs S4), confirming that the proposed DFL framework delivers consistent cost reductions irrespective of the uncertainty source to which it is applied. 

The relative impact of having two uncertainty sources is most directly revealed by comparing S1 and S2, while isolating a single source of uncertainty. 
Load uncertainty alone (S2) incurs a total cost of 1,813.97 \texteuro \,, exceeding the PV-only counterpart (S1, 1,774.18 \texteuro \,) by 2.19\% and producing roughly three times the regret (3.39\% vs 1.12\%). Notably, the nature of the cost degradation also differs: load uncertainty in S2 generates negligible shedding (0.126 kWh) but higher grid exchange costs, whereas PV uncertainty in S1 triggers more frequent shedding events (3.595 kWh) yet at a lower overall cost.

Under the most practically relevant conditions, where both load and PV generation are simultaneously uncertain, the proposed DFL framework demonstrates a clear advantage over the conventional MSE approach. The fully DFL-trained configuration (S5) reduces total operational cost by 71.40 \texteuro \, (3.78\%) relative to the MSE-MSE baseline (S8), with regret falling from 7.53\% to 3.46\%. A decomposition of this improvement through the hybrid configurations reveals a pronounced asymmetry: substituting only the load forecaster with its DFL counterpart (S6) captures nearly the entire cost reduction (1,817.53 \texteuro \, vs 1,815.26 \texteuro \, for S5), whereas substituting only the PV forecaster (S7) yields a more modest improvement (1,854.00 \texteuro \,). This asymmetry is consistent with the single-source analysis of Section 6.1 and is further reflected in the load shedding volumes, which drop from 20.4 kWh under S8 to effectively zero under S5 and S6. The near-complete elimination of shedding events suggests that the DFL-trained load forecaster internalizes the asymmetric penalty structure of the EMS, learning to produce forecasts that systematically avoid the underestimation errors responsible for costly real-time curtailment.

\begin{figure*}[!t]
\centering
\includegraphics[width=\textwidth]{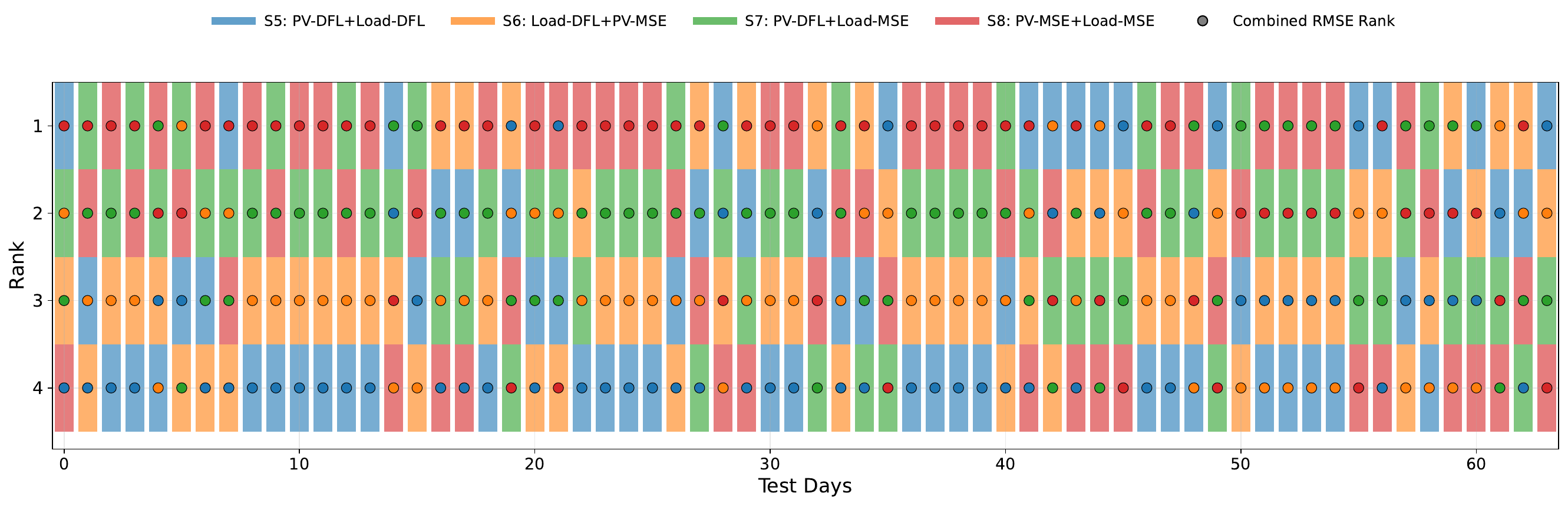}
\caption{Daily ranking of the four combined-forecaster scenarios over the test set (40\,kW unconstrained case). Colored bars indicate the daily operational cost rank, while black-edged scatter markers denote the combined forecast RMSE rank ($\sqrt{\mathrm{RMSE}_{\mathrm{load}}^2 + \mathrm{RMSE}_{\mathrm{PV}}^2}$). Rank\,1 corresponds to the best-performing scenario on each day. Divergence between bars and markers reveals days where forecast accuracy does not align with economic performance.}
\label{fig:rank_40kw}
\end{figure*}

The daily ranking analysis presented in Figure \ref{fig:rank_40kw} provides further insight into the relationship between forecast accuracy and operational cost. The comparison between daily cost rankings and combined RMSE rankings shows that forecast accuracy and operational cost frequently disagree: the most accurate configuration on a given day is often not the cheapest. This decoupling is a direct consequence of the non-linear relationship between forecast error and operational cost imposed by the two-stage EMS, where the direction and timing of errors matter more than their magnitude.

\begin{figure*}[!t]
\centering
\includegraphics[width=\textwidth]{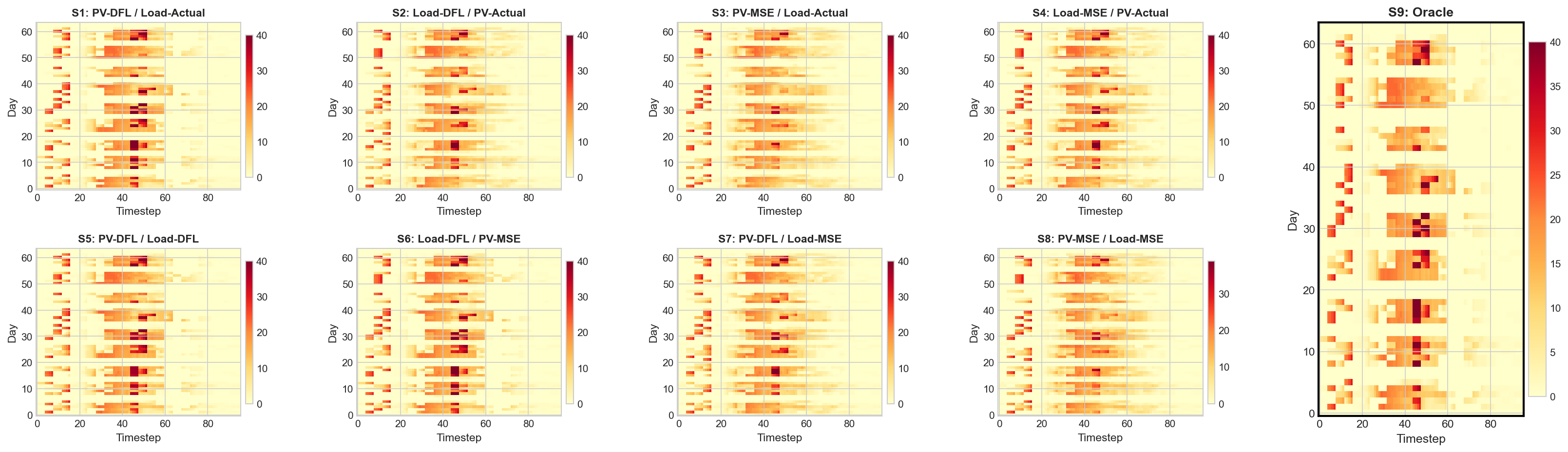}
\caption{Day-by-timestep heatmap of grid purchase power across all forecasting scenarios (40\,kW grid). Top row: single-source uncertainty scenarios (S1--S4); bottom row: combined forecaster scenarios (S5--S8). The oracle (S9, right panel) provides the perfect-information dispatch reference. Each cell represents the grid import at a given 15-min timestep on a given test day.}
\label{fig:heatmap_pbuy}
\end{figure*}

\begin{figure*}[!t]
\centering
\includegraphics[width=\textwidth]{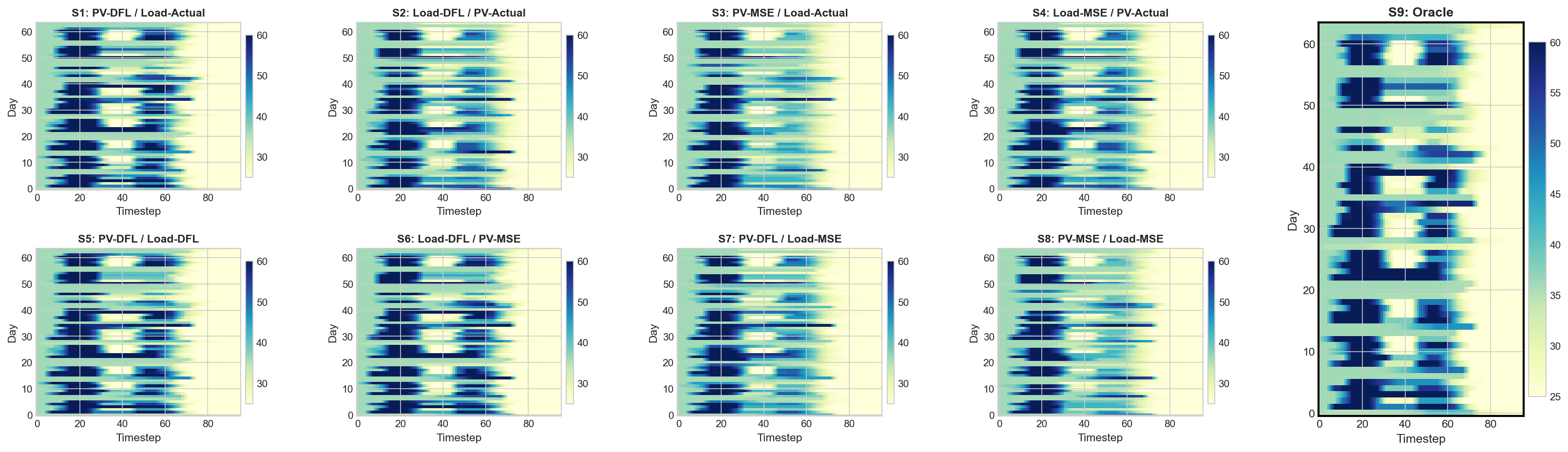}
\caption{Day-by-timestep heatmap of BESS SOE across all forecasting scenarios (40\,kW grid). Top row: single-source uncertainty scenarios (S1--S4); bottom row: combined forecaster scenarios (S5--S8). The oracle (S9, right panel) shows the optimal storage trajectory under perfect information. Each cell represents the stored energy at a given 15-min timestep on a given test day.}
\label{fig:heatmap_soe}
\end{figure*}

The operational heatmaps support the preceding analysis at the dispatch level. DFL scenarios produce grid purchase patterns (Figure \ref{fig:heatmap_pbuy}) and BESS state of energy trajectories (Figure \ref{fig:heatmap_soe}) that more closely resemble the oracle baseline, with reduced import saturation and more anticipatory battery cycling. These differences, while moderate under the 40 kW scenario due to the available grid flexibility, foreshadow the more pronounced operational divergence observed under constrained conditions in Section 6.2.

\begin{figure*}[!t]
\centering
\begin{minipage}[t]{0.49\textwidth}
    \centering
    \includegraphics[width=\textwidth]{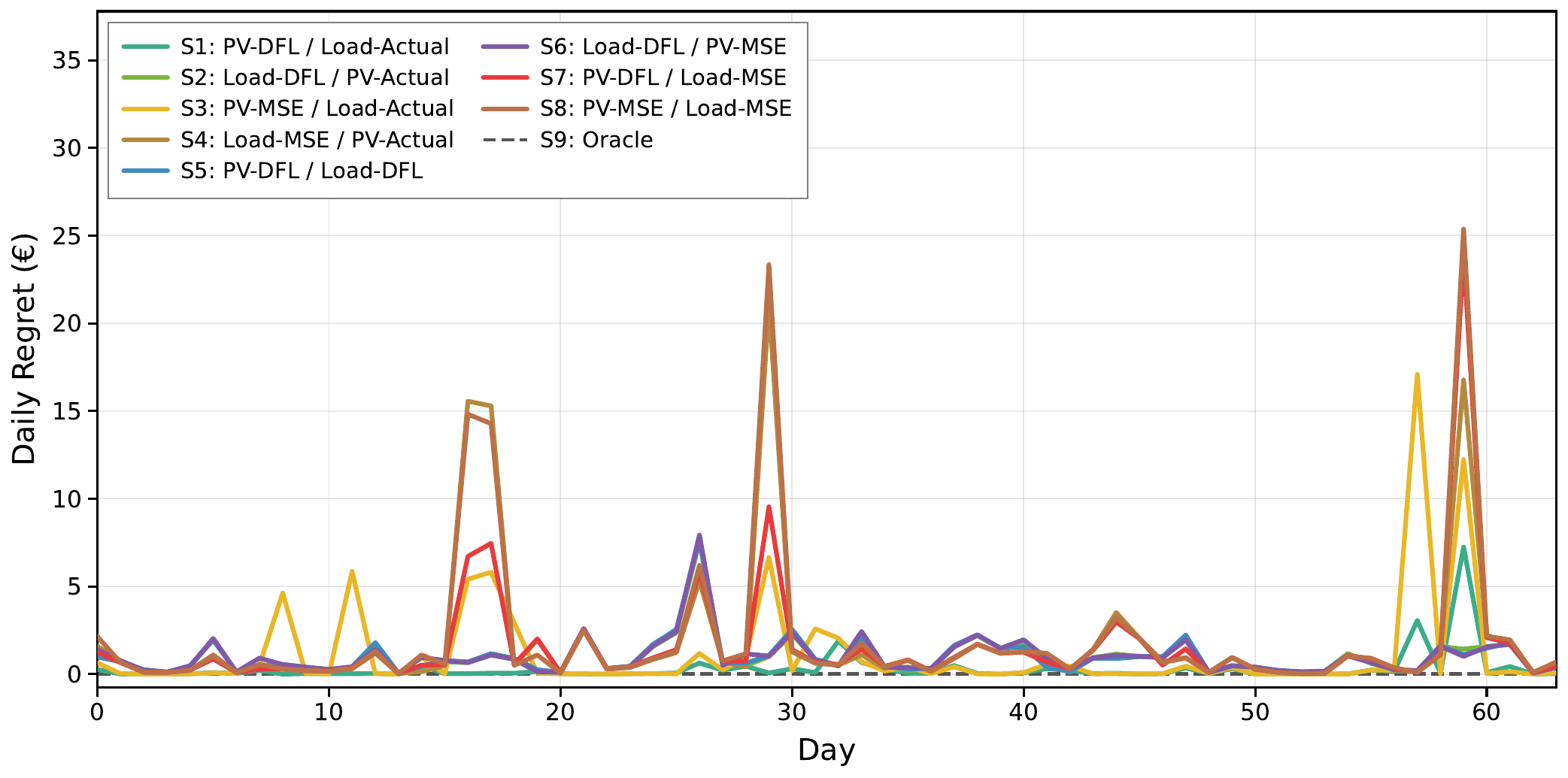}
\end{minipage}
\hfill
\begin{minipage}[t]{0.49\textwidth}
    \centering
    \includegraphics[width=\textwidth]{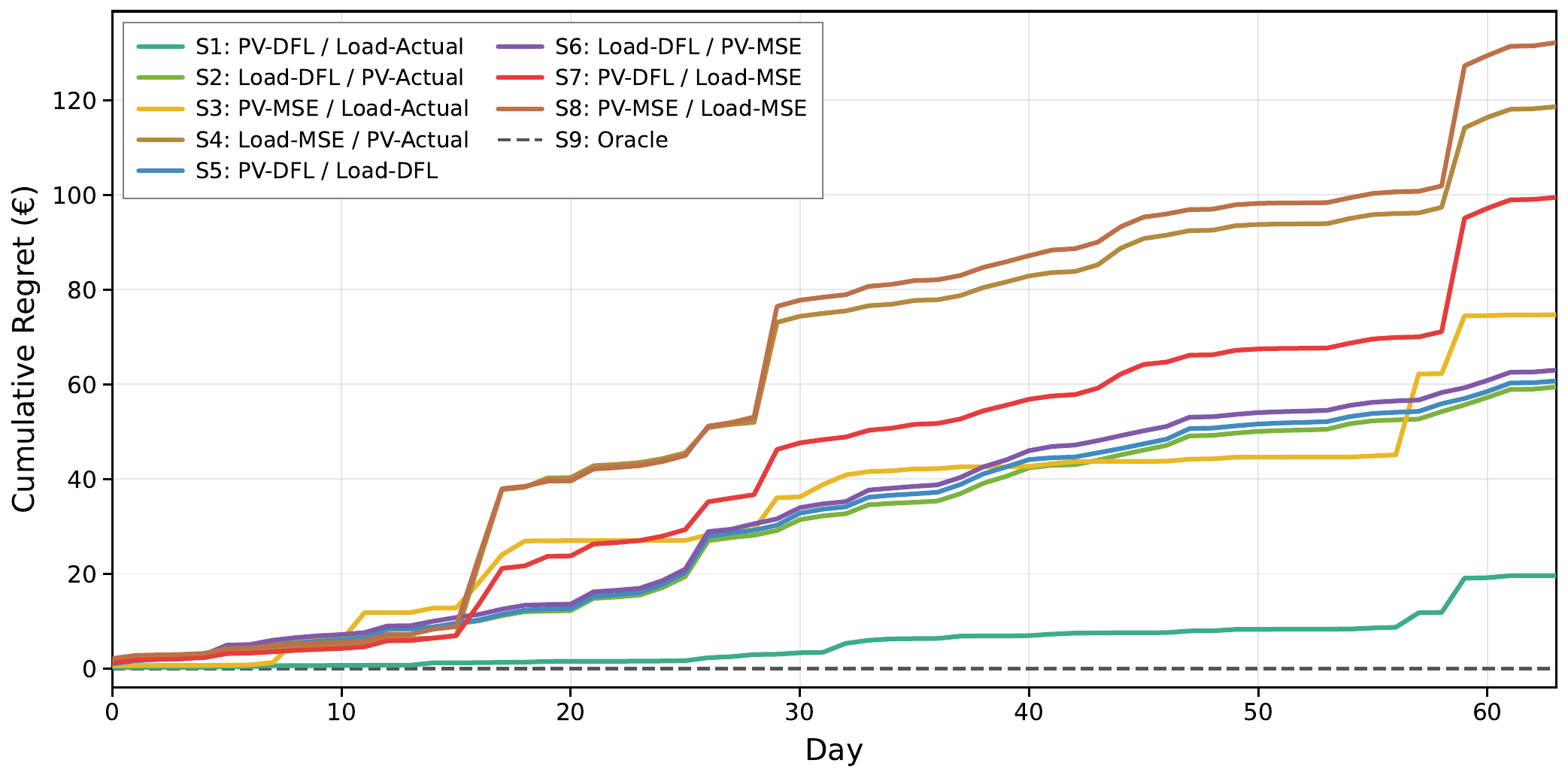}
\end{minipage}
\caption{Daily regret (left) and cumulative regret (right) of all forecasting scenarios relative to the oracle baseline over the test set (40\,kW grid). The dashed line represents the oracle with perfect information.}
\label{fig:regret_40kw}
\end{figure*}

Finally, an interesting observation emerges from the daily win rate analysis. Under the unconstrained 40 kW scenario, where grid flexibility can absorb most forecast-induced inefficiencies, the MSE-MSE baseline (S8) occasionally achieves the lowest daily cost. This occurs predominantly on days with low demand variability, where the statistically unbiased MSE forecast aligns well with realized conditions. Yet despite this, the DFL configurations achieve a lower total cost over the 64-day horizon as it is demonstrated in Figure \ref{fig:regret_40kw}. In this sense, the DFL framework functions as an economic risk mitigation mechanism, trading small cost concessions on favorable days for substantial savings on adverse ones.

\begin{table*}[!t]
\caption{Operational performance of eight forecasting scenarios and the oracle lower bound for BESS scheduling under the DFL framework (40\,kW grid connection). Scenarios are ranked by total cost within each group.}
\label{tab:results_40kw}
\centering
\small
\setlength{\tabcolsep}{3.6pt}
\renewcommand{\arraystretch}{1.2}
\begin{tabular}{@{}c cc cc cc c cc cc c@{}}
\toprule
\textbf{ID} & \multicolumn{2}{c}{\textbf{Scenario}} & \multicolumn{2}{c}{\textbf{Load Error}} & \multicolumn{2}{c}{\textbf{PV Error}} & \textbf{Total Cost} & \multicolumn{2}{c}{\textbf{Regret}} & \multicolumn{2}{c}{\textbf{Load Shedding}} & \textbf{Win Rate} \\
\cmidrule(lr){2-3} \cmidrule(lr){4-5} \cmidrule(lr){6-7} \cmidrule(lr){9-10} \cmidrule(lr){11-12}
 & \textbf{Load} & \textbf{PV} & \textbf{MAE} & \textbf{RMSE} & \textbf{MAE} & \textbf{RMSE} & & & & & & \\
 & \textbf{Input} & \textbf{Input} & \textbf{(kW)} & \textbf{(kW)} & \textbf{(kW)} & \textbf{(kW)} & \textbf{(\euro{})} & \textbf{(\euro{})} & \textbf{(\%)} & \textbf{(kWh)} & \textbf{(\%)} & \textbf{(\%)} \\
\midrule
\multicolumn{13}{@{}l}{\textit{Single-Source Uncertainty Scenarios}} \\
\addlinespace[2pt]
S1 & Actual & DFL & -- & -- & 0.495 & 1.087 & \textbf{1774.18} & 19.65 & 1.12 & 3.595 & 0.0471 & 45.31 \\
S2 & DFL & Actual & 2.408 & 4.425 & -- & -- & 1813.97 & 59.45 & 3.39 & 0.126 & 0.0016 & 3.12 \\
S3 & Actual & MSE & -- & -- & 0.471 & 1.037 & 1829.24 & 74.71 & 4.26 & 17.956 & 0.2350 & 45.31 \\
S4 & MSE & Actual & 1.945 & 3.730 & -- & -- & 1873.14 & 118.62 & 6.76 & 17.896 & 0.2342 & 3.12 \\
\addlinespace[5pt]
\multicolumn{13}{@{}l}{\textit{Combined Forecasters Scenarios}} \\
\addlinespace[2pt]
S5 & DFL & DFL & 2.408 & 4.425 & 0.495 & 1.087 & \textbf{1815.26} & 60.74 & 3.46 & 0.000 & 0.0000 & 0.00 \\
S6 & DFL & MSE & 2.408 & 4.425 & 0.471 & 1.037 & 1817.53 & 63.01 & 3.59 & 0.000 & 0.0000 & 1.56 \\
S7 & MSE & DFL & 1.945 & 3.730 & 0.495 & 1.087 & 1854.00 & 99.48 & 5.67 & 12.267 & 0.1605 & 0.00 \\
S8 & MSE & MSE & 1.945 & 3.730 & 0.471 & 1.037 & 1886.66 & 132.14 & 7.53 & 20.367 & 0.2665 & 1.56 \\
\addlinespace[5pt]
\midrule
S9 & Actual & Actual & -- & -- & -- & -- & 1754.52 & 0.00 & 0.00 & 0.000 & 0.0000 & -- \\
\bottomrule
\end{tabular}

\vspace{4pt}
{\footnotesize
\begin{flushleft}
Regret is defined as $\Delta C = C_{\mathrm{s}} - C_{\mathrm{oracle}}$.
Load Shedding (\%) represents the share of total energy curtailed with respect to the aggregate load consumption across the test set.
Win Rate denotes the fraction of test days on which a scenario achieves the minimum daily operating cost.
\end{flushleft}}
\end{table*}

\subsection{Case 2 Scenario}
\label{sec:case2}
The second case study reduces the grid connection capacity to 18 kW import and export, effectively eliminating the grid's role as a real-time buffer against suboptimal day-ahead commitments. Under these conditions, the oracle configuration (S9) achieves a total cost of 1,771.68 \texteuro \, (Table \ref{tab:results_18kw}), representing an increase of only 17.16 \texteuro (0.98\%) relative to the unconstrained oracle. This near-equivalence is significant: it demonstrates that the grid constraint itself imposes negligible additional cost under perfect information. Consequently, the substantial cost increases observed in the forecasted configurations throughout this section are entirely attributable to forecast errors and their propagation through the two-stage EMS, rather than to the physical constraint per se.

The single-source uncertainty scenarios (S1–S4) reveal a dramatic amplification of the DFL advantage under constrained grid conditions. On the PV side, the DFL-trained forecaster reduces operational cost by 26.01\% (631.06 \texteuro \,) relative to its MSE counterpart (S1 vs S3), compared to 3.01\% in Case 1. On the load side, the corresponding improvement rises from 3.16\% to 25.69\% (685.23 \texteuro \,, S2 vs S4), exhibiting a similar amplification factor. The load shedding volumes provide the most direct explanation for this cost escalation: the MSE load forecaster (S4) incurs 228.3 kWh of shedding over the test horizon, while its DFL counterpart (S2) reduces this to 39.4 kWh. On the PV side, the contrast is even starker, with the MSE configuration (S3) producing 175.3 kWh of shedding compared to effectively zero for the DFL configuration (S1). Under the constrained 18 kW grid connection, the system can no longer compensate for forecast underestimation through increased grid import; consequently, every underestimation event propagates directly into the unserved energy penalty cost.

The combined forecaster results under constrained conditions represent the most practically significant finding of this study. The DFL-DFL configuration (S5) reduces total operational cost by 875.02 \texteuro \, (30.60\%) relative to MSE-MSE (S8), with regret falling from 61.40\% to 12.01\% and shedding volume decreasing by an order of magnitude, from 279.4 kWh to 33.5 kWh. While the load forecaster remains the dominant driver of this improvement, a notable departure from Case 1 emerges in the behavior of the hybrid configurations. In the unconstrained scenario, S6 achieved a cost nearly identical to S5, suggesting that PV forecast quality had minimal operational impact. Under constrained conditions, however, S6 incurs 2,087.11 \texteuro \,, a 102.64 \texteuro \, premium over S5, indicating that PV forecast errors now carry tangible cost consequences that the system can no longer absorb. The reverse configuration (S7) at 2,389.85 \texteuro \, confirms the expected hierarchy: decision-focused load forecasting remains the more consequential improvement, but the gap between S5 and S6 demonstrates that extending DFL to PV forecasting provides a meaningful incremental benefit that becomes increasingly relevant as grid flexibility diminishes.

\begin{figure*}[!t]
\centering
\includegraphics[width=\textwidth]{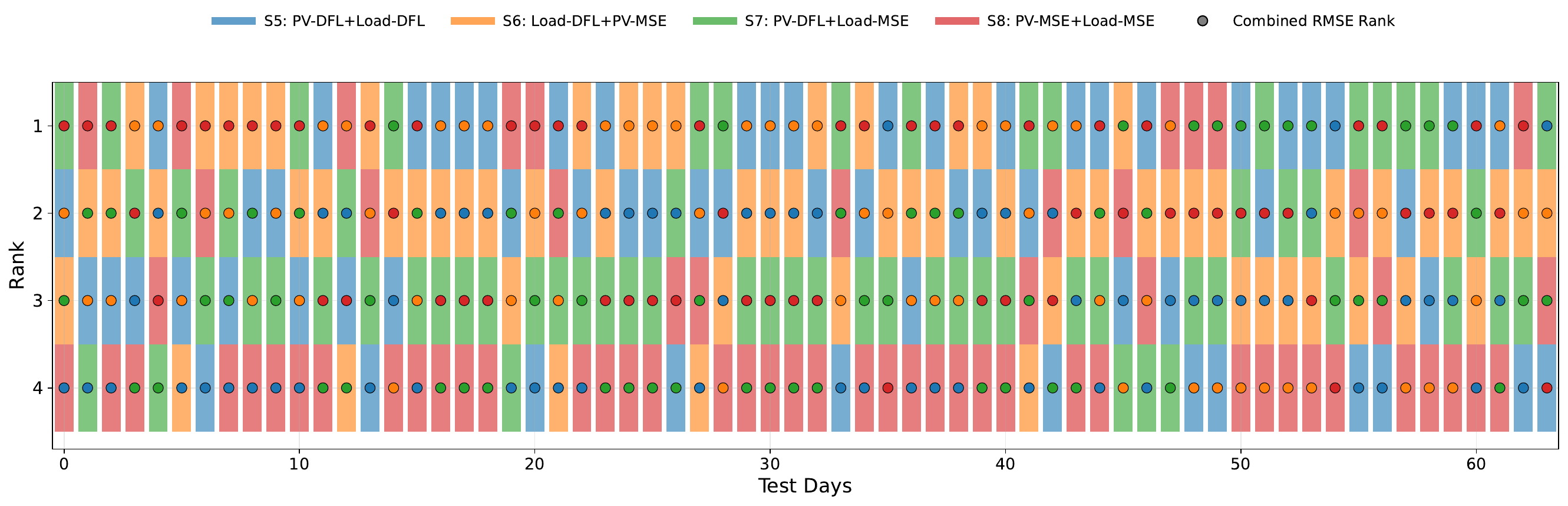}
\caption{Daily ranking of the four combined-forecaster scenarios over the test set (18\,kW grid). Colored bars indicate the daily operational cost rank, while black-edged scatter markers denote the combined forecast RMSE rank ($\sqrt{\mathrm{RMSE}_{\mathrm{load}}^2 + \mathrm{RMSE}_{\mathrm{PV}}^2}$). Rank\,1 corresponds to the best-performing scenario on each day. Divergence between bars and markers reveals days where forecast accuracy does not align with economic performance.}
\label{fig:rank_18kw}
\end{figure*}

The daily ranking analysis (Figure \ref{fig:rank_18kw}) confirms that the DFL advantage under constrained conditions is both larger in magnitude and more consistent across the test horizon. The DFL configurations (S5, S6) now dominate ranks 1–2 with substantially greater regularity than in Case 1, while the MSE-MSE baseline (S8) is relegated to the last rank on the majority of test days. The divergence between the combined RMSE rank and the daily cost rank is notably more pronounced than in the unconstrained scenario, reinforcing that forecast accuracy becomes an increasingly unreliable proxy for economic performance as grid flexibility decreases. Furthermore, the DFL configurations now consistently surpass the MSE-MSE baseline in terms of daily win rate, indicating that the constrained operating environment leaves insufficient margin for accuracy-optimized forecasters to compete even on benign days.

\begin{figure*}[!t]
\centering
\includegraphics[width=\textwidth]{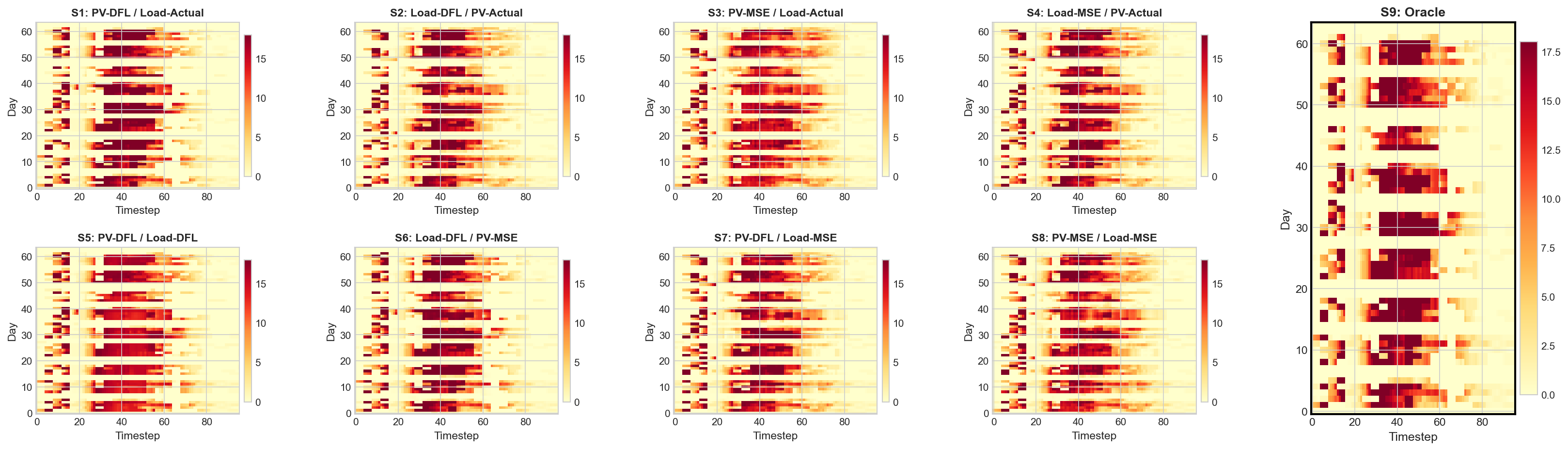}
\caption{Day-by-timestep heatmap of grid purchase power across all scenarios (18\,kW grid). Top row: single-source uncertainty (S1--S4); bottom row: combined forecasters (S5--S8). The oracle (S9, right panel) provides the perfect-information dispatch reference. Each cell represents the grid import at a given 15-min timestep on a given test day.}
\label{fig:heatmap_pbuy_18kw}
\end{figure*}

\begin{figure*}[h!]
\centering
\includegraphics[width=\textwidth]{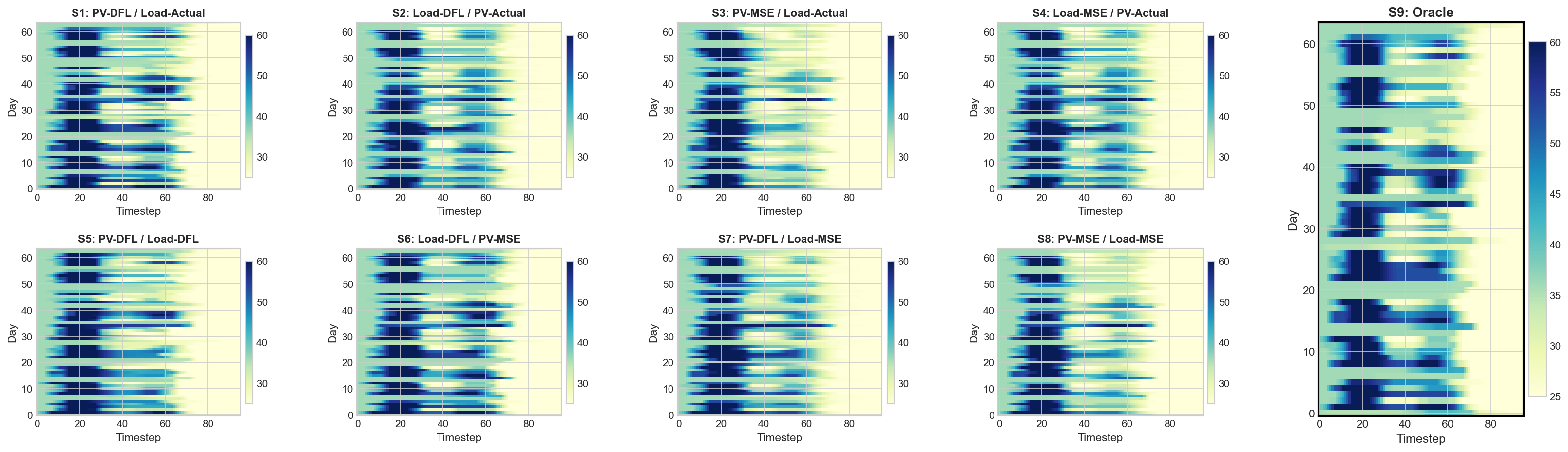}
\caption{Day-by-timestep heatmap of BESS SOE across all scenarios (18\,kW grid). Top row: single-source uncertainty (S1--S4); bottom row: combined forecasters (S5--S8). The oracle (S9, right panel) shows the optimal storage trajectory under perfect information. Each cell represents the stored energy at a given 15-min timestep on a given test day.}
\label{fig:heatmap_soe_18kw}
\end{figure*}

The operational heatmaps tell a considerably starker story than in Case~1. The grid purchase patterns (Figure~\ref{fig:heatmap_pbuy_18kw}) show MSE scenarios persistently saturating the 18~kW import ceiling, while DFL configurations maintain modulated import profiles closer to the oracle. The BESS SOE trajectories (Figure~\ref{fig:heatmap_soe_18kw}) reveal the underlying mechanism: MSE configurations frequently deplete the battery to $SOE_{\min}$, exhausting the system's only controllable reserve, whereas DFL scenarios preserve higher SOE levels during critical periods in a manner consistent with the oracle's anticipatory storage strategy. The visual contrast between DFL and MSE heatmaps is substantially more pronounced than under unconstrained conditions, confirming that the cost amplification in Table~\ref{tab:results_18kw} reflects fundamentally different dispatch behaviors. The following cross-case analysis quantifies this amplification systematically.

\begin{figure*}[!t]
\centering
\begin{minipage}[t]{0.49\textwidth}
    \centering
    \includegraphics[width=\textwidth]{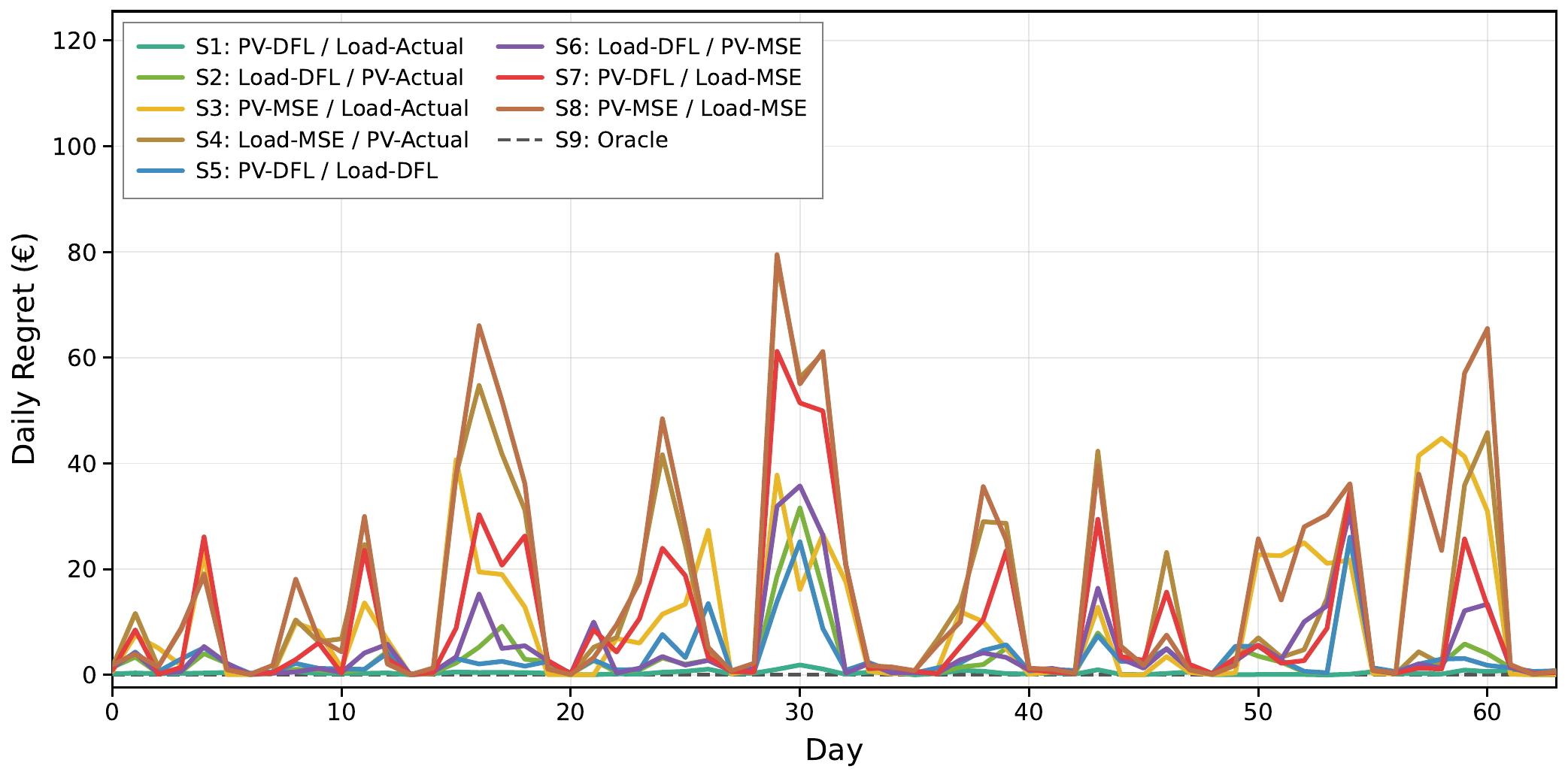}
\end{minipage}
\hfill
\begin{minipage}[t]{0.49\textwidth}
    \centering
    \includegraphics[width=\textwidth]{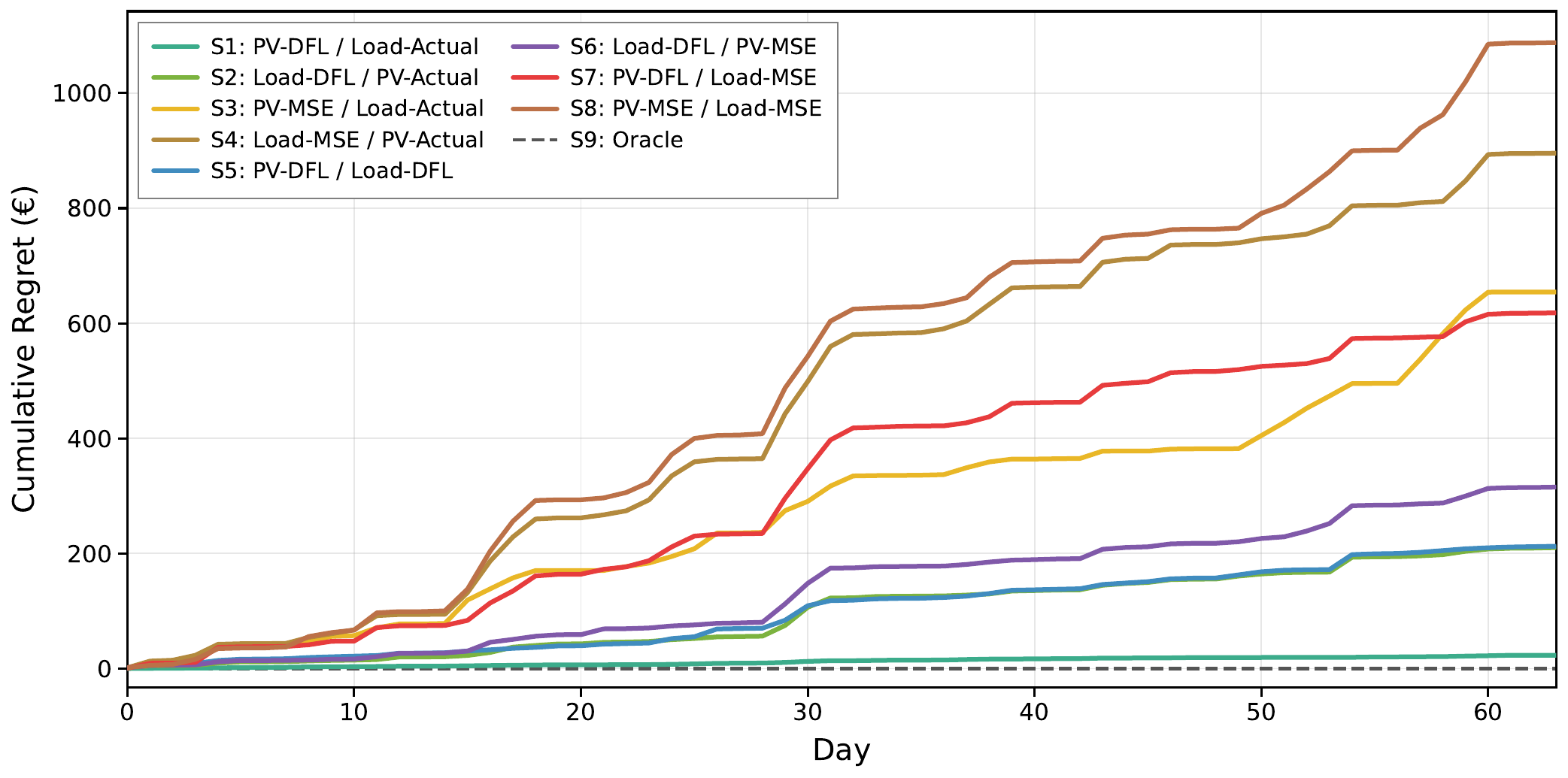}
\end{minipage}
\caption{Daily regret (left) and cumulative regret (right) of all forecasting scenarios relative to the oracle baseline over the test set (18\,kW grid). The dashed line represents the oracle with perfect information.}
\label{fig:regret_18kw}
\end{figure*}

The cumulative regret trajectories (Figure \ref{fig:regret_18kw}, right panel) reflect this dominance quantitatively: MSE-based scenarios accumulate regret at an accelerating rate, diverging sharply from the DFL trajectories which remain comparatively flat throughout the test period. The daily regret profile (Figure \ref{fig:regret_18kw}, left panel) reveals the mechanism behind this divergence, MSE configurations exhibit frequent high-regret spikes reaching 70–80 \texteuro \, per day, driven by severe shedding events that are largely absent from the DFL scenarios. The operational insurance mechanism identified in Case 1, where the DFL forecaster traded marginal optimality on benign days for protection against costly outcomes, is no longer a secondary effect; under constrained conditions, it becomes the dominant driver of cost separation between the two methodologies.

\begin{table*}[!t]
\caption{Operational performance of eight forecasting scenarios and the oracle lower bound for BESS scheduling under the DFL framework (18\,kW grid). Scenarios are ranked by total cost within each group.}
\label{tab:results_18kw}
\centering
\small
\setlength{\tabcolsep}{3.6pt}
\renewcommand{\arraystretch}{1.2}
\begin{tabular}{@{}c cc cc cc c cc cc c@{}}
\toprule
\textbf{ID} & \multicolumn{2}{c}{\textbf{Scenario}} & \multicolumn{2}{c}{\textbf{Load Error}} & \multicolumn{2}{c}{\textbf{PV Error}} & \textbf{Total Cost} & \multicolumn{2}{c}{\textbf{Regret}} & \multicolumn{2}{c}{\textbf{Load Shedding}} & \textbf{Win Rate} \\
\cmidrule(lr){2-3} \cmidrule(lr){4-5} \cmidrule(lr){6-7} \cmidrule(lr){9-10} \cmidrule(lr){11-12}
 & \textbf{Load} & \textbf{PV} & \textbf{MAE} & \textbf{RMSE} & \textbf{MAE} & \textbf{RMSE} & & & & & & \\
 & \textbf{Input} & \textbf{Input} & \textbf{(kW)} & \textbf{(kW)} & \textbf{(kW)} & \textbf{(kW)} & \textbf{(\euro{})} & \textbf{(\euro{})} & \textbf{(\%)} & \textbf{(kWh)} & \textbf{(\%)} & \textbf{(\%)} \\
\midrule
\multicolumn{13}{@{}l}{\textit{Single-Source Uncertainty Scenarios}} \\
\addlinespace[2pt]
S1 & Actual & DFL & -- & -- & 0.738 & 1.598 & \textbf{1795.02} & 23.34 & 1.32 & 0.000 & 0.0000 & 68.75 \\
S2 & DFL & Actual & 2.101 & 4.007 & -- & -- & 1982.06 & 210.38 & 11.87 & 39.447 & 0.5162 & 0.00 \\
S3 & Actual & MSE & -- & -- & 0.471 & 1.037 & 2426.08 & 654.40 & 36.94 & 175.262 & 2.2934 & 25.00 \\
S4 & MSE & Actual & 1.945 & 3.730 & -- & -- & 2667.29 & 895.62 & 50.55 & 228.274 & 2.9871 & 1.56 \\
\addlinespace[5pt]
\multicolumn{13}{@{}l}{\textit{Combined Forecasters Scenarios}} \\
\addlinespace[2pt]
S5 & DFL & DFL & 2.101 & 4.007 & 0.738 & 1.598 & \textbf{1984.47} & 212.79 & 12.01 & 33.513 & 0.4385 & 0.00 \\
S6 & DFL & MSE & 2.101 & 4.007 & 0.471 & 1.037 & 2087.11 & 315.43 & 17.80 & 68.424 & 0.8954 & 1.56 \\
S7 & MSE & DFL & 1.945 & 3.730 & 0.738 & 1.598 & 2389.85 & 618.18 & 34.89 & 151.437 & 1.9817 & 3.12 \\
S8 & MSE & MSE & 1.945 & 3.730 & 0.471 & 1.037 & 2859.49 & 1087.81 & 61.40 & 279.372 & 3.6558 & 0.00 \\
\addlinespace[5pt]
\midrule
S9 & Actual & Actual & -- & -- & -- & -- & 1771.68 & 0.00 & 0.00 & 0.000 & 0.0000 & -- \\
\bottomrule
\end{tabular}

\vspace{6pt}
\noindent\footnotesize
Regret is defined as $\Delta C = C_{\mathrm{s}} - C_{\mathrm{oracle}}$.
Load Shedding (\%) represents the share of total energy curtailed with respect to the aggregate load consumption across the test set.
Win Rate denotes the fraction of test days on which a scenario achieves the minimum daily operating cost.
\end{table*}

\subsection{Discussion}
\label{sec:discussion}
The results presented in Sections~\ref{sec:case1} and~\ref{sec:case2} raise three interrelated questions: why does a statistically inferior forecaster produce economically superior decisions, what governs the magnitude of this advantage, and what are the practical implications for microgrid operators? The following discussion addresses each in turn.

The central finding of this work is that forecast accuracy and operational cost do not share a monotonic relationship in microgrid energy management systems. In decision-dependent settings, the economic impact of a forecast error is not determined by its absolute magnitude alone but depends on its direction, timing, and interaction with exogenous parameters such as grid capacity, electricity prices, and PV availability. Replacing the traditional PTO paradigm, in which the forecaster is trained independently of the downstream decision problem, with a decision-focused approach yields forecasters that are statistically inferior yet economically superior. This non-monotonicity is clearly evidenced in the experimental results. In Case~1, the DFL-DFL configuration (S5) exhibits a load MAE of 2.408~kW and a PV MAE of 0.495~kW, both higher than the MSE-MSE baseline (S8, 1.945~kW and 0.471~kW respectively), yet achieves a total operational cost 71.40 \texteuro \, lower. In Case~2, the accuracy gap persists (load MAE: 2.101 vs 1.945~kW; PV MAE: 0.738 vs 0.471~kW) while the cost gap widens dramatically to 875.02 \texteuro \,. These results demonstrate that optimizing for statistical accuracy not only fails to minimize operational cost but can actively work against it when the cost landscape is asymmetric.

The cross-case comparison further reveals that the magnitude of the DFL advantage is governed by the degree of operational flexibility available to the microgrid. Under normal grid conditions (Case~1), the utility grid connection acts as a low-cost buffer that absorbs forecast errors through additional grid import at market price, limiting the operational consequences of suboptimal BESS commitments. When this buffer is reduced to 18~kW (Case~2), forecast errors that were previously compensated through grid exchange now propagate directly into load shedding events. This transition is quantified in Table~\ref{tab:dfl_vs_mse_paired}: the DFL cost advantage over MSE increases from 3.78\% to 30.60\% in the joint forecaster comparison (S5 vs S8), while shedding volumes under the MSE\nobreakdash-MSE configuration rise from 20.4~kWh to 279.4~kWh. The DFL forecaster, trained end\nobreakdash-to\nobreakdash-end through this penalty structure, learns a conservative bias that systematically avoids the high\nobreakdash-cost region of the decision space by overestimating load during critical periods, thereby pre\nobreakdash-positioning the BESS to meet demand peaks without reliance on grid import. 

These results carry a broader methodological implication: the DFL framework achieves risk-averse operational behavior as an emergent property of end-to-end training, without recourse to the explicit uncertainty modeling required by conventional stochastic or robust optimization approaches. Scenario-based methods require the construction of representative uncertainty sets and scale poorly with the number of stochastic variables, while minimax formulations tend to produce overly conservative solutions driven by worst-case assumptions that rarely materialize. The proposed DFL approach circumvents both limitations by allowing the forecaster to internalize the cost asymmetry directly from the training signal, producing forecasts that are implicitly calibrated to the operational risk profile of the system. This simplification comes at the cost of a single structural requirement: the optimization problem must remain convex to enable gradient computation through the KKT conditions.

In summary, the discussion reveals that the proposed DFL framework offers a fundamentally different approach to forecasting in microgrid energy management: one that prioritizes decision quality over statistical accuracy. The experimental evidence across two operational scenarios, eight forecasting configurations, and 64 test days consistently demonstrates that end-to-end training through the optimization layer produces forecasters that are better aligned with the true operational objective. The magnitude of this alignment benefit is modest under unconstrained conditions but becomes substantial as grid flexibility decreases, suggesting that the value proposition of DFL is strongest precisely in the operating regimes where accurate scheduling matters most.

\begin{table*}[ht!]
\caption{Paired DFL vs MSE comparison across three forecasting configurations and both case studies.}
\label{tab:dfl_vs_mse_paired}
\centering
\setlength{\tabcolsep}{4pt}
\renewcommand{\arraystretch}{1.2}
\begin{tabular}{@{}l l l cc cc cc c@{}}
\toprule
 & & & \multicolumn{2}{c}{\textbf{DFL Cost Reduction}} & \multicolumn{2}{c}{\textbf{Regret vs Oracle (\%)}} & \multicolumn{2}{c}{\textbf{Shedding (kWh)}} & \textbf{DFL Win} \\
\cmidrule(lr){4-5} \cmidrule(lr){6-7} \cmidrule(lr){8-9}
\textbf{IDs} & \textbf{Comparison} & \textbf{Case} & \textbf{(\euro{})} & \textbf{(\%)} & \textbf{DFL} & \textbf{MSE} & \textbf{DFL} & \textbf{MSE} & \textbf{Rate (\%)} \\
\midrule
\multirow{2}{*}{S1 vs S3} & \multirow{2}{*}{PV Forecaster} & 40\,kW & 55.06 & 3.01 & 1.12 & 4.26 & 3.595 & 17.956 & 54.7 \\
 & & 18\,kW & 631.06 & 26.01 & 1.32 & 36.94 & 0 & 175.262 & 75.0 \\
\addlinespace[4pt]
\multirow{2}{*}{S2 vs S4} & \multirow{2}{*}{Load Forecaster} & 40\,kW & 59.17 & 3.16 & 3.39 & 6.76 & 0.126 & 17.896 & 32.8 \\
 & & 18\,kW & 685.23 & 25.69 & 11.87 & 50.55 & 39.447 & 228.274 & 73.4 \\
\addlinespace[4pt]
\multirow{2}{*}{S5 vs S8} & \multirow{2}{*}{Joint Forecaster} & 40\,kW & 71.40 & 3.78 & 3.46 & 7.53 & 0 & 20.367 & 37.5 \\
 & & 18\,kW & 875.02 & 30.60 & 12.01 & 61.40 & 33.513 & 279.372 & 71.9 \\
\addlinespace[4pt]
\bottomrule
\end{tabular}

\vspace{6pt}
\noindent\footnotesize
Cost Reduction $= C_{\mathrm{MSE}} - C_{\mathrm{DFL}}$\,; positive values indicate DFL outperforms MSE.
Regret $= (C_{\mathrm{s}} - C_{\mathrm{oracle}}) / C_{\mathrm{oracle}} \times 100$.
Win Rate denotes the fraction of days DFL achieves a lower daily cost than its MSE counterpart.
\end{table*}

\section{Conclusions}
\label{sec:conclusions}

The conventional PTO paradigm trains forecasters to minimize statistical error independently of the downstream scheduling problem, disregarding the fact that forecast errors carry asymmetric and time-varying operational consequences. This paper addresses this gap by proposing a DFL framework that trains GRU forecasters for both load demand and PV generation directly on the operational cost of a two-stage microgrid EMS, with gradients backpropagated through the optimization layer via KKT-based implicit differentiation. Validated on the MG\textsuperscript{2}Lab experimental testbed across two grid cases and eight forecasting scenarios, the proposed framework reduces operational cost by 3.78\% under normal conditions and by 30.60\% under grid-constrained conditions relative to the MSE-trained baseline, while nearly eliminating load shedding events. These results confirm that forecast accuracy and decision quality do not share a monotonic relationship, and that the value of decision-focused training amplifies as operational flexibility decreases.

Despite the positive results, some limitations should be acknowledged. First, the framework employs a single-shot day-ahead commitment without intraday re-adjustment; integrating a rolling-horizon model predictive control scheme could further improve performance. Second, the two forecasters are trained independently under perfect foresight of the other uncertainty source, and a simultaneous training procedure represents a natural extension. Last, while the outperforming results of DFL a interesting aspect would be the comparison of training procedure time and the applicability in edge devices.


\bibliographystyle{unsrtnat}
\bibliography{references}

\end{document}